\documentclass[aps, preprint,nofootinbib]{revtex4}
\usepackage{graphicx}

\begin{document}
\global\long\def\met{\not\!\! E_{T}}
\global\long\def\fl{\mathcal{F}_{L}}
\global\long\def\fr{\mathcal{F}_{R}}
\global\long\def\dvtb{\delta V_{tb}}
\global\long\def\wtb{Wtb}
\global\long\def\wtbp{W^{+}t\bar{b}}
\global\long\def\ztt{Zt\bar{t}}
\global\long\def\zttr{Zt_{R}\bar{t}_{R}}
\global\long\def\zttl{Zt_{L}\bar{t}_{L}}
\global\long\def\zbb{Zb\bar{b}}
\global\long\def\zbbl{Zb_{L}\bar{b}_{L}}
\global\long\def\zbbr{Zb_{R}\bar{b}_{R}}

\preprint{ANL-HEP-PR-09-48, EFI-09-19}
\vspace*{-0.5cm}
\title{Model Independent Constraints Among the $Wtb$, $Zb{\bar b}$, and $Zt{\bar t}$ Couplings}
\vspace*{0.5cm}
\author{\vspace{0.5cm} Edmond L. Berger}
\email{berger@anl.gov}
\affiliation{High Energy Physics Division, Argonne National Laboratory, Argonne, IL 60439}
\author{Qing-Hong Cao}
\email{caoq@hep.anl.gov}
\affiliation{High Energy Physics Division, Argonne National Laboratory, Argonne, IL 60439 \\
Enrico Fermi Institute, University of Chicago, Chicago, IL 60637}
\author{Ian Low}
\email{ilow@anl.gov}
\affiliation{High Energy Physics Division, Argonne National Laboratory, Argonne, IL 60439\\
\mbox{Department of Physics \& Astronomy, Northwestern University, Evanston, IL 60208}
\vspace{1cm}}

\begin{abstract}
Using an effective Lagrangian approach, we perform a model-independent analysis
of the interactions among electroweak gauge bosons and the third generation
quarks, {\it i.e.} the $Wtb$, $Z t \bar{t}$ and $Z b \bar{b}$ couplings.   
After one imposes the known experimental constraint on the $Z b_L b_L$ coupling, 
we show that the electroweak $SU(2)_L\times U(1)_Y$ symmetry of the standard model
specifies the pattern of deviations of the $Z t_L t_L$ and $W t_L b_L$ couplings,
independent of underlying new physics scenarios.
We study implications of the predicted pattern with data on the 
single top quark and $Z t \bar{t}$ associated production processes
at the Large Hadron Collider. Such an analysis could in principle allow for 
a determination of the $Wtb$ coupling without prior knowledge of $|V_{tb}|$, 
which is otherwise difficult to achieve.
\end{abstract}

\maketitle

%
%
\section{Introduction}

It is widely accepted that interactions of the third generation quarks,
the top and bottom quarks, offer a window into possible new physics
beyond the standard model (SM) of particle physics. Since the top
quark mass is close to the Fermi scale 
$v\equiv(\sqrt{2} G_{F})^{-1/2}\approx 246~\rm{GeV}$,
its interactions are thought to be sensitive to the mechanism
of electroweak symmetry breaking. Indeed, in the majority of models
beyond the SM, deviations are predicted from the SM in the top quark
interactions, especially in their couplings with the electroweak gauge
bosons, such as the $Wtb$, $Zb\bar{b}$, and $Zt\bar{t}$ couplings.
Interactions of the top quark have yet to be measured precisely, allowing
possible room for deviations from the SM. Experimental constraints
on the bottom and top quark couplings with the $W$ and $Z$ bosons
provide an interesting contrast: the bottom quark left-handed coupling
$Zb_{L}b_{L}$ is determined precisely by the measurements at LEP
I and II~\cite{Abdallah:2008yx}, whereas the $Zt_{L}t_{L}$ coupling
is virtually not measured so far. The $Wt_{L}b_{L}$ coupling was
confirmed only recently in single top quark production at the Fermilab
Tevatron~\cite{Abazov:2006gd,Abazov:2009ii,Aaltonen:2009jj}. It
will be probed further at the CERN Large Hadron Collider (LHC).

In this work we investigate the range of possible deviations of the
couplings of the top and bottom quarks with the electroweak gauge
bosons by exploiting the contrasting experimental constraints mentioned
above. Adopting an effective Lagrangian approach \cite{Weinberg:1978kz},
we parametrize the effects of new physics in terms of higher dimensional
operators constructed from SM fields: $c_{i}{\cal O}_{i}/\Lambda^{n-4}$,
where $n$ is the mass dimension of the operator ${\cal O}_{i}$,
$\Lambda$ is the scale of new physics, and $c_{i}$ is a numerical
coefficient assumed to be of order unity unless otherwise specified.
Such an approach is valid when all new particles are heavier than
the Fermi scale and whose effects can be integrated out of the effective
theory, as we assume. However, this assumption does not preclude the
possibility that new particles at the TeV scale could be produced
directly and observed at the LHC.

Within the general context in which we work, we demonstrate that the
$SU(2)_{L}\times U(1)_{Y}$ symmetry of the SM yields correlations
among the possible deviations of the couplings $Zb_{L}b_{L}$, $Zt_{L}t_{L}$,
and $Wt_{L}b_{L}$. Once the stringent experimental bound on $Zb_{L}b_{L}$
is taken into account, a unique prediction follows on the size of
the $Zt_{L}t_{L}$ and $Wt_{L}b_{L}$ couplings. The prediction is
a striking manifestation of $SU(2)_{L}\times U(1)_{Y}$ invariance,
and it is independent of the underlying new physics at the electroweak
scale.

After an exposition of the general operator analysis in Section II,
we use existing experimental constraints to show that deviations of
the $Wtb$ and $Zt{\bar{t}}$ couplings from their SM values can depend
on only two parameters, $\fl$ and $\fr$, and we present the allowed
ranges of these parameters for different new physics cutoff scales
and Higgs boson masses. In Section III, we survey the landscape of new
physics models which may be expected to modify these couplings. The
prospects that $\fl$ and $\fr$ can be determined better from data
from the LHC are addressed in Section IV where we focus on single top
quark and $\ztt$ associated production.
Our results from Section II show deviations from the SM of
the single top quark total and differential cross sections depend
only on $\fl$, whereas the cross section for $\ztt$ associated production
is influenced by both $\fl$ and $\fr$.

The presence of contributions from two operators in $\ztt$ associated
production means that observation, or at least a bound on the effects
of new physics in this process is in principle possible provided one
looks sufficiently differentially into the distributions of particles
in the final state. We began our study expecting to show that spin
correlations in the final state would offer significant advantages,
particularly those between the top quark spin and the decay lepton
from $t\to bW^{+}(\to\ell^{+}\nu)$. However, as we conclude from
a detailed simulation in Sec.~IV, experimental cuts distort the most
telling distributions and appear to leave a sample of effectively
unpolarized top quarks. This result, while disappointing, does not
appear to have been established previously. Turning adversity to advantage,
we remark that the difficulty in finding evidence for new physics
in the $\ztt$ final state may be interpreted as good news for new
physics searches. The associated production of $\ztt$ is a dominant
background for new physics models~\cite{Meade:2006dw,Cao:2006wk}.
If a new physics model predicts the production of highly polarized
top quark pairs, then one might be able to see evidence above the effectively
unpolarized dominant background. Among the final state distributions
in $\ztt$ associated production, we confirm that the opening angle
between the two charged leptons from the $Z$ boson decay, 
$\Delta\phi(\ell^{\prime+}\ell^{\prime-})$,
appears useful for limiting the couplings $\fl$ and $\fr$, as is
pointed out in Ref.~\cite{Baur:2004uw}.

We also comment that in single top quark production $\fl$ is always multiplied by
the CKM matrix element $|V_{tb}|$, which is very close to one in the SM.
Therefore if both $\fl$ and $\fr$ could be extracted from $\ztt$ production 
through the total and differential cross-sections, one could infer the value of
$|V_{tb}|$ from measurements in singlet top quark production.

We present the results of our analysis as a set of two-dimensional
plots of the deviation from the SM of the associated production cross
section $\delta\sigma_{\ztt}$ versus the deviation of the single
top quark cross section $\delta\sigma_{t}$ for integrated luminosities
of $300\,{\rm fb}^{-1}$ and $3000\,{\rm fb}^{-1}$ at the LHC. These
expectations are contrasted with estimates for a high energy electron-positron
linear collider (LC).  A comparison is also shown with predictions based
on recent models of new physics including a top-prime model~\cite{Dobrescu:2009vz},
a right-handed $t'$ model~\cite{Chivukula:1998wd},
and a model with sequential fourth generation quarks that mix with 
the third generation~\cite{Kribs:2007nz}.  
Our analysis suggests that 
about $3000\,{\rm fb}^{-1}$ at the LHC would be needed to improve model 
independent constraints on  $\fl$ and $\fr$ to the same level achievable at 
a 500 GeV LC with $100\,{\rm fb}^{-1}$ of integrated luminosity.  On the 
other hand, $300\,{\rm fb}^{-1}$ at the LHC would be sufficient to delineate 
which new physics models could be possible.  

In addition to the study in Ref.~\cite{Baur:2004uw} of $\ztt$ associated
production mentioned above, other earlier work related to ours includes
the study in Ref.~\cite{Batra:2006iq}  of the prospects for
measuring the $Wtb$ coupling at a linear collider as a test of different
models of new physics. The authors investigate $t\bar{t}$ production
and single top quark production, as we do, but they do not focus on
the correlations.  

%
%
\section{Operator Analysis and Existing Constraints}

We begin with a general assumption that effects beyond the SM are described by a set
of higher dimensional operators made out of the SM fields only.
Once the (approximate) symmetry of the SM is assumed, these operators
start at dimension six~\cite{Weinberg:1978kz}.
A complete list of operators is presented in~\cite{Buchmuller:1985jz},
whose notation we follow.  Therefore the effective Lagrangian we work with
is of the form 
\begin{equation}
\mathcal{L}_{eff}=\mathcal{L}_{SM}
+ \frac{1}{\Lambda^{2}}\sum_{i}\left(c_{i}\mathcal{O}_{i}+h.c.\right)
+ {\cal O}\left(\frac{1}{\Lambda^{3}}\right),
\label{eq:eft}
\end{equation}
where the coefficients $c_{i}$'s are numerical constants 
parameterizing the strength of the
non-standard interactions. The excellent agreement between the SM
expectations and data indicates that deviations from the SM are small.
Hence, when computing the effects of new operators we can restrict
ourselves to the interference terms between $\mathcal{L}_{SM}$ and
the operators $\mathcal{O}_{i}$, i.e. we can work to first order
in the coefficients $c_{i}$.

\subsection{Operator Analysis}\label{sect:opana}

Three types of dimension-six operators contribute
to the $Wtb$, $Zb{\bar{b}}$, and $Zt{\bar{t}}$ couplings: 1) operators
involving scalars and vectors, 2) operators involving fermions and
vectors, and 3) operators involving vectors, fermions, and scalars.
We discuss all three types in turn in the following.

Operators involving scalars and vectors enter through the self-energy
of the electroweak gauge boson: 
${\cal O}_{\phi W}=\frac{1}{2}(\phi^{\dagger}\phi)\, W_{\mu\nu}^{I}W_{\mu\nu}^{I}$,
${\cal O}_{\phi B}=\frac{1}{2}(\phi^{\dagger}\phi)\, B_{\mu\nu}B_{\mu\nu}$,
and 
${\cal O}_{WB}=(\phi^{\dagger}\tau^{I}\phi)\, W_{\mu\nu}^{I}B_{\mu\nu}$,
where $\phi$ denotes the SM scalar Higgs doublet, $W_{\mu\nu}^{I}$
and $B_{\mu\nu}$ are the field-strength tensors for the $SU(2)_{L}$
and $U(1)_{Y}$ gauge bosons respectively, and $\tau^{I}=\sigma^{I}/2$
is the usual $SU(2)_{L}$ generator in the fundamental representation.
Operators ${\cal O}_{\phi W}$ and ${\cal O}_{\phi B}$ arise only
after new physics is integrated out at the loop level~\cite{Arzt:1994gp}.
Their corrections to the self-energy are of order
$(1/16\pi^{2})\times(v^{2}/\Lambda^{2})$,
where $v$ is the vacuum expectation value of the Higgs. 
Operator ${\cal O}_{WB}$ can be generated by tree-level exchange
of new particles, but its contribution is related to the $S$ parameter.
It is highly constrained by precision electroweak data, which requires
$c_{WB}\sim{\cal O}(10^{-2})$ ~\cite{Barbieri:2004qk} for $\Lambda\sim1$
TeV. Thus, all three operators are effectively suppressed by a loop
factor for new physics at about the 1~TeV scale, and we neglect them
in this study.

Operators of the second type necessarily have two fermions carrying
one gauge-covariant derivative and one field strength~\cite{Buchmuller:1985jz},
such as ${\cal O}_{qW}=i(\bar{q}\tau^{I}\gamma_{\mu}D_{\nu}q)\, W_{\mu\nu}^{I}$.
These operators give a correction of order $p^{2}/\Lambda^{2}$ to
the couplings of interest here, where $p$ is the typical momentum
scale in the process and can be taken to be the Fermi scale $p\sim v$.
However, such operators correspond to vertices with only three legs.
They appear only once new physics is integrated out at the loop level%
~\cite{Arzt:1994gp,delAguila:2000aa,delAguila:2000rc}.
Their natural size is again of order $(1/16\pi^{2})\times(v^{2}/\Lambda^{2})$.
A similar conclusion can be obtained from naive dimensional analysis~\cite{Manohar:1983md},
since each derivative in the operator carries an extra $4\pi$ suppression.
Therefore, we do not consider operators of the second type here.

Operators of the third type can be generated both at the tree-level and at the 
loop level. The loop-induced operators, such as
$\bar{q} \sigma^{\mu\nu} \tau^I b_R \phi W_{\mu\nu}^I$,
$\bar{q} \sigma^{\mu\nu} \tau^I t_R \tilde{\phi} W_{\mu\nu}^I$,
$\bar{q} \sigma^{\mu\nu} t_R \tilde{\phi} B_{\mu\nu}$, 
and $\bar{q} \sigma^{\mu\nu} b_R \phi B_{\mu\nu}$,
are not included in our analysis as they are suppressed by small coefficients
of order $1/16\pi^2$~\cite{Arzt:1994gp}. We focus our attention on 
the tree-level induced operators of the third type throughout this paper. 

The dimension-six operators of the third type are \begin{eqnarray}
\mathcal{O}_{\phi q}^{(1)} & = & i\left(\phi^{\dagger}D_{\mu}\phi\right)\left(\bar{q}\gamma^{\mu}q\right),\\
\mathcal{O}_{\phi q}^{(3)} & = & i\left(\phi^{\dagger}\tau^{I}D_{\mu}\phi\right)\left(\bar{q}\gamma^{\mu}\tau^{I}q\right),\\
\mathcal{O}_{\phi t} & = & i\left(\phi^{\dagger}D_{\mu}\phi\right)\left(\bar{t}_{R}\gamma^{\mu}t_{R}\right),\\
\mathcal{O}_{\phi b} & = & i\left(\phi^{\dagger}D_{\mu}\phi\right)\left(\bar{b}_{R}\gamma^{\mu}b_{R}\right),\\
\mathcal{O}_{\phi\phi} & = & \left(\phi^{\dagger}\epsilon D_{\mu}\phi\right)\left(\bar{t}_{R}\gamma^{\mu}b_{R}\right),\end{eqnarray}
where $D_{\mu}$ is the gauge-covariant derivative; $q$ is the left-handed
top-bottom $SU(2)_{L}$ doublet $q=(t_{L},\, b_{L})$; $t_{R}(b_{R})$
are the corresponding right-handed isosinglets; and $\epsilon=i\sigma^{2}$
is the two-dimensional antisymmetric tensor. Corrections from these
operators are of order $v^{2}/\Lambda^{2}$ and could be generated
from integrating out new physics at the tree-level. 
For example, the operator $\mathcal{O}_{\phi q}^{(3)}$ can be induced 
by a heavy $t^{\prime}$ quark mixing with the top quark, which is present 
in many theories beyond the SM.

It is worth mentioning that equations of motion can be used to turn
the ${\cal O}_{WB}$ operator into type 3) operators that are universal
in flavors~\cite{Strumia:1999jm}. Thus, the statement we are about
to make applies to contributions from ${\cal O}_{WB}$ as well.

Upon symmetry breaking $\langle\phi\rangle=v/\sqrt{2}$. 
The set of operators of the third type generate the following corrections to
the couplings $Wtb$, $\ztt$ and $\zbb$: \begin{eqnarray}
\mathcal{O}_{\wtb} & = & \frac{c_{\phi q}^{(3)}v^{2}}{\Lambda^{2}}\frac{g_{2}}{\sqrt{2}}W_{\mu}^{+}\bar{t}_{L}\gamma^{\mu}b_{L}-\frac{c_{\phi\phi}v^{2}}{2\Lambda^{2}}\frac{g_{2}}{\sqrt{2}}W_{\mu}^{+}\bar{t}_{R}\gamma^{\mu}b_{R}+h.c.,\label{eq:wtb}\\
\mathcal{O}_{\ztt} & = & \frac{\left(c_{\phi q}^{(3)}-c_{\phi q}^{(1)}\right)v^{2}}{\Lambda^{2}}\frac{\sqrt{g_{1}^{2}+g_{2}^{2}}}{2}Z_{\mu}\bar{t}_{L}\gamma^{\mu}t_{L}-\frac{c_{\phi t}v^{2}}{\Lambda^{2}}\frac{\sqrt{g_{1}^{2}+g_{2}^{2}}}{2}Z_{\mu}\bar{t}_{R}\gamma^{\mu}t_{R},\label{eq:ztt}\\
\mathcal{O}_{\zbb} & = & -\frac{\left(c_{\phi q}^{(1)}+c_{\phi q}^{(3)}\right)v^{2}}{\Lambda^{2}}\frac{\sqrt{g_{1}^{2}+g_{2}^{2}}}{2}Z_{\mu}\bar{b}_{L}\gamma^{\mu}b_{L}-\frac{c_{\phi b}v^{2}}{\Lambda^{2}}\frac{\sqrt{g_{1}^{2}+g_{2}^{2}}}{2}Z_{\mu}\bar{b}_{R}\gamma^{\mu}b_{R},\label{eq:zbb}\end{eqnarray}
where $g_{2}$ and $g_{1}$ are the coupling strengths of the $SU(2)$
and $U(1)$ gauge interaction, respectively.

%
%

\subsection{Existing Constraints}\label{subsect:constraint}

Among the five operators listed above, $\mathcal{O}_{\phi\phi}$ is
tightly constrained by recent data on the rare decay of $b\to s\gamma$,
$-0.0007<\frac{c_{\phi\phi}v^{2}}{2\Lambda^{2}}<0.0025$~\cite{Chetyrkin:1996vx,
Larios:1999au,Burdman:1999fw,Grzadkowski:2008mf},
provided there is no accidental cancellation with contributions from
other new physics effects, such as those produced by the four-fermion
operator $b\bar{s}t\bar{t}$.~\footnote{This operator can be generated, for example, 
by exchanging a heavy $W'$ vector boson.} 
As a result, effects from $\mathcal{O}_{\phi\phi}$ are small and are not considered further.

For the purpose of our analysis, the most useful experimental constraints
on the five operators of the third type are the precise measurements
of $R_{b}$ and $A_{FB}^{(b)}$ at LEP II~\cite{Abdallah:2008yx}. 
These bound the $\zbb$ coupling. The measured value of the $\zbbl$ coupling agrees with
the SM prediction at the 0.25\% level. It enforces the relation 
\begin{equation}
c_{\phi q}^{(3)}+c_{\phi q}^{(1)} \simeq 0,
\label{eq:cancel}
\end{equation}
which, in turn, implies that the deviations in the $Wt_{L}b_{L}$
and $Zt_{L}t_{L}$ couplings are controlled by the same parameter
$c_{\phi q}^{(3)}\simeq-c_{\phi q}^{(1)}$. In other words, the $SU(2)_{L}\times U(1)_{Y}$
symmetry of the SM predicts a certain pattern in the deviations of
the electroweak gauge boson couplings to the third generation quarks
that is independent of the possible new physics beyond the SM.

Certain subgroups of the custodial symmetry~\cite{Sikivie:1980hm}
which protect the $\rho (\equiv m_W/m_Z \cos\theta_W)$ parameter 
can also preserve the $Zb_{L}b_{L}$
coupling in the SM~\cite{Agashe:2006at}, resulting in $c_{\phi q}^{(3)}+c_{\phi q}^{(1)}=0$
exactly. This result is obtained if the top and bottom quarks are
embedded in suitable representations of $SU(2)_{L}\times SU(2)_{R}$
whose diagonal group $SU(2)_{V}$ serves as the custodial symmetry. If
one implements the custodial symmetry in this way, the $Zt_{R}t_{R}$
coupling is also protected, $c_{\phi t}=0$~\cite{Agashe:2006at}.
In this work we are not concerned with the underlying reason for the
smallness in the deviation in the $Zb_{L}b_{L}$ coupling. We take
Eq.~(\ref{eq:cancel}) as an empirical statement, allowing us to
be model-independent. A recent study of top compositeness and the
third generation couplings to electroweak gauge bosons within the
the framework of Ref.~\cite{Agashe:2006at} can be found in Ref.~\cite{Pomarol:2008bh}.

After Eq.~(\ref{eq:cancel}) is imposed, the $Wtb$ and $Zt\bar{t}$
couplings depend on only two unknown parameters: 
\begin{eqnarray}
\mathcal{O}_{\wtb} & = & \frac{g}{\sqrt{2}}\,\fl\, W^{+}\bar{t}_{L}\gamma^{\mu}b_{L}+h.c.\ ,\\
\mathcal{O}_{\ztt} & = & \frac{g}{2c_{w}}Z_{\mu}\left(2\fl\,\bar{t}_{L}\gamma^{\mu}t_{L}+\fr\,\bar{t}_{R}\gamma^{\mu}t_{R}\right),
\end{eqnarray}
where $\fl\equiv c_{\phi q}^{(3)}v^{2}/\Lambda^{2}$, $\fr\equiv-c_{\phi t}v^{2}/\Lambda^{2}$,
and $c_{w}=\cos\theta_{w}$ is the cosine of the Weinberg angle. Notice
the relation between the coefficients of the left-handed neutral and
charged currents%
~\footnote{A similar relation is pointed out in Ref.~\cite{Carlson:1994bg} 
in the context of an electroweak chiral Lagrangian, even though
the implication of such a relation was not studied.}: 
\begin{equation}
g_{\ztt}^{L}=2g_{\wtb}^{L}=2\fl.\label{eq:CC-NC}
\end{equation}
This equation states the pattern of deviations predicted by the electroweak
symmetry of the SM, after the stringent constraint on $Zb_{L}b_{L}$ is imposed. 
Notice further that the right-handed coupling $\zbbr$
does not enter the stated correlation in the $Wtb$ and $Zt{\bar{t}}$
couplings, leaving room for the interesting possibility that $b_{R}$
could be (partially) composite. As it turns out, a positive shift
of the $\zbbr$ coupling, say $\delta g_{\zbbr}\simeq+0.02$, would
explain the $3\sigma$ deviation in the forward-backward asymmetry
$A_{FB}^{(b)}$ measured by the LEP and SLAC Large Detector  experiments~\cite{aleph:2005dia}.

Within the low-energy effective theory, $\fl$ and $\fr$ in ${\cal O}_{Ztt}$
induce one-loop corrections to the $\rho$ parameter and the $\zbb$
vertex, which are associated with the observables $\epsilon_{1}$
and $\epsilon_{b}$ summarized in Ref.~\cite{Altarelli:2004mr}.
The pure SM one-loop contributions to $\epsilon_{1}$ and $\epsilon_{b}$
are~\cite{Altarelli:1997et} 
\begin{eqnarray}
\epsilon_{1}^{SM} & = & \frac{3G_{F}m_{t}^{2}}{4\sqrt{2}\pi^{2}}-\frac{3G_{F}m_{W}^{2}}{4\sqrt{2}\pi^{2}}\tan^{2}\theta_{w}\log\left(\frac{m_{H}}{m_{Z}}\right),\label{eq:epsilon1_sm}\\
\epsilon_{b}^{SM} & = & -\frac{G_{F}m_{t}^{2}}{4\sqrt{2}\pi^{2}},
\label{eq:epsilonb_sm}
\end{eqnarray}
where $m_H (m_W, m_Z, m_t)$ denotes the mass of Higgs boson ($W$-boson, $Z$-boson, top quark), respectively. 
The contributions from the anomalous couplings $\fl$ and $\fr$ are~\cite{Larios:1999au}
\begin{eqnarray}
\delta\epsilon_{1} & = & \frac{3m_{t}^{2}G_{F}}{2\sqrt{2}\pi^{2}}\left(\fr-\fl\right)\ln\left(\frac{\Lambda^{2}}{m_{t}^{2}}\right),\label{eq:epsilon1}\\
\delta\epsilon_{b} & = & \frac{m_{t}^{2}G_{F}}{2\sqrt{2}\pi^{2}}\left(2\fl-\frac{1}{4}\fr\right)\ln\left(\frac{\Lambda^{2}}{m_{t}^{2}}\right),\label{eq:epsilonb}
\end{eqnarray}
where $\Lambda$ is the cutoff of the low-energy effective theory
and is taken to be the scale of new physics. Notice that $\epsilon_{1}^{SM}$ depends on $m_{H}$, 
which implies that the allowed region of $\fl$ and $\fr$ will depend on $m_{H}$ as well as on $\Lambda$.

\begin{figure}
\includegraphics[clip,scale=0.3]{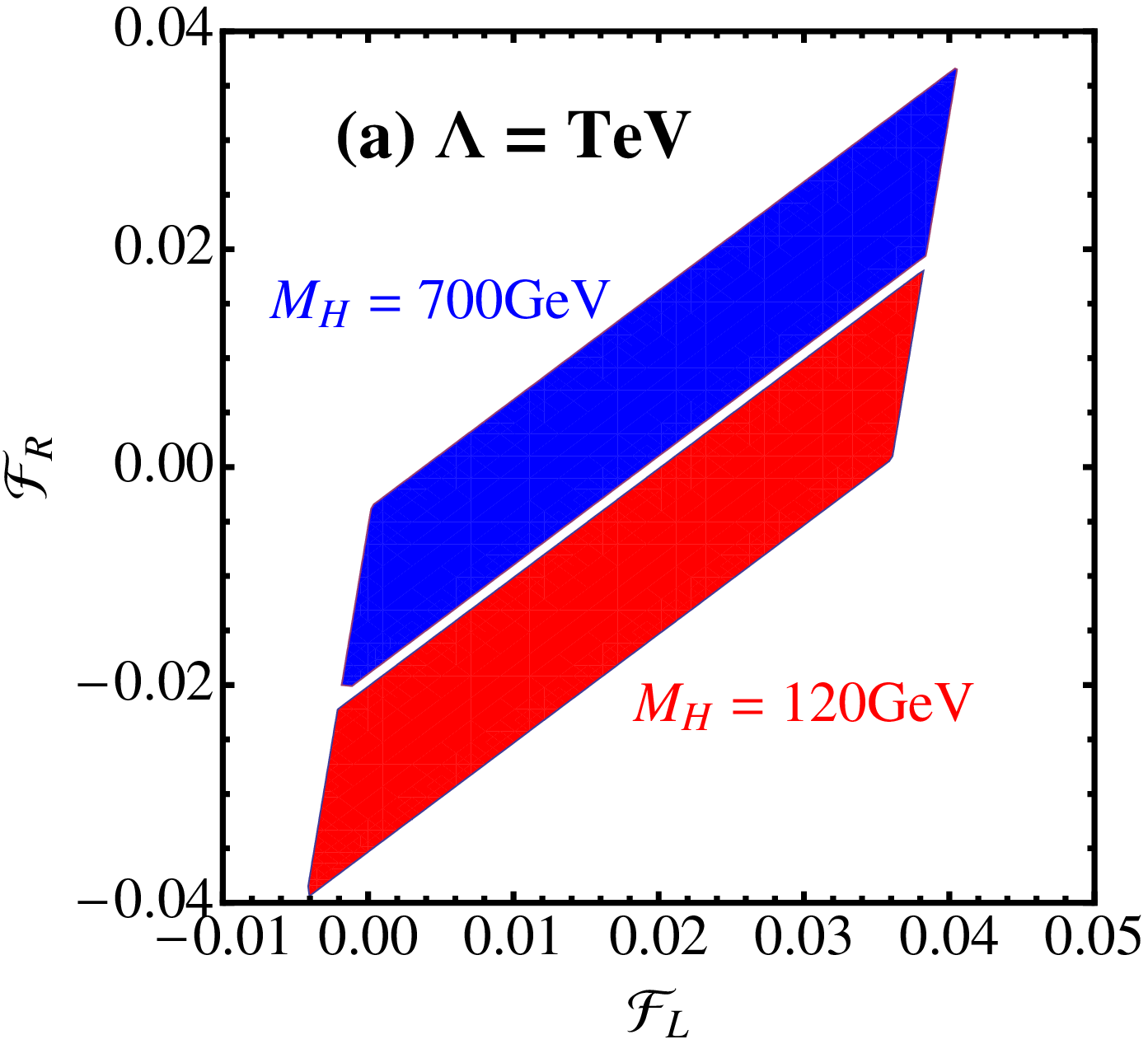}
\includegraphics[clip,scale=0.3]{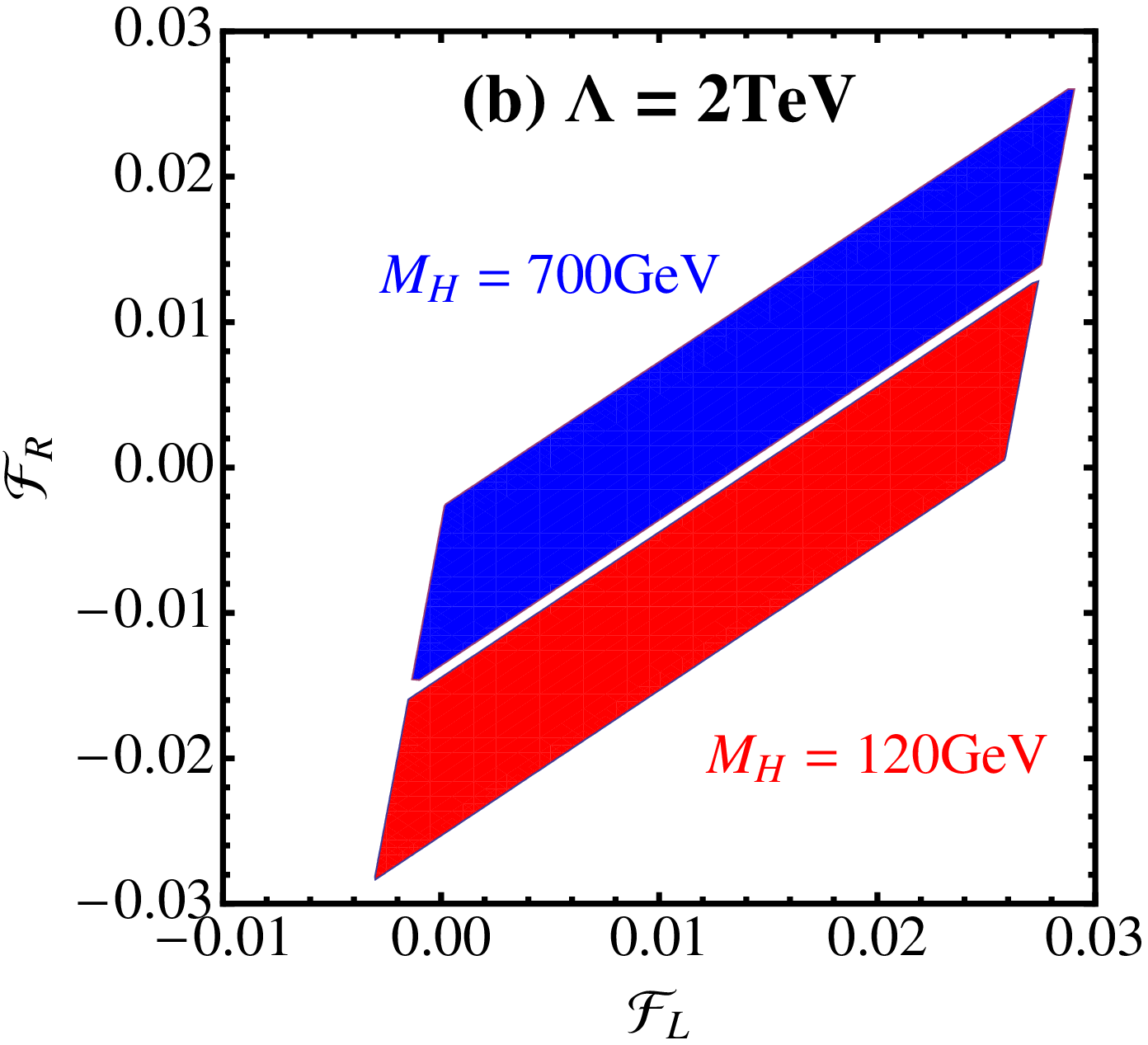}
\includegraphics[clip,scale=0.3]{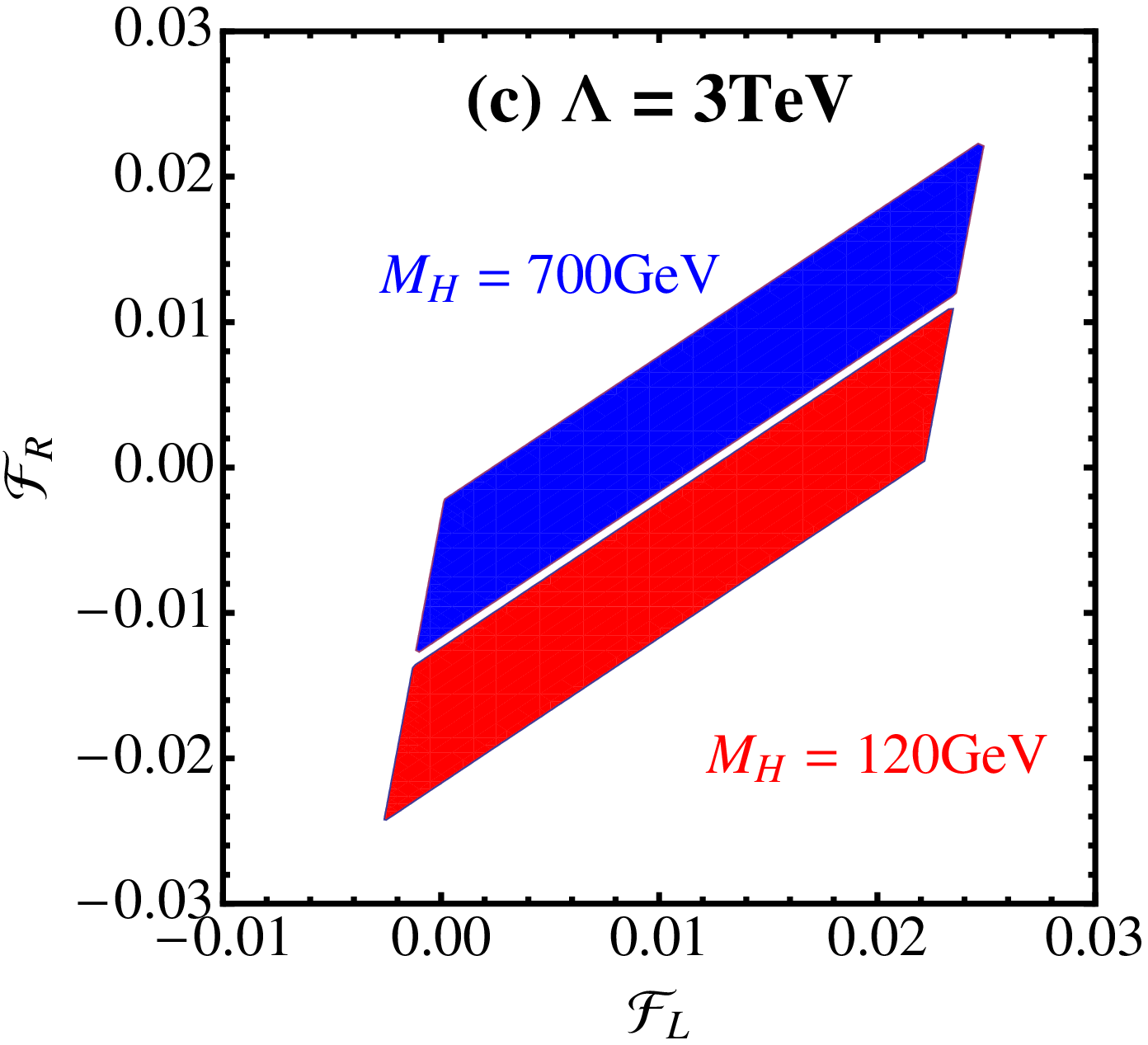}\\
\includegraphics[clip,scale=0.3]{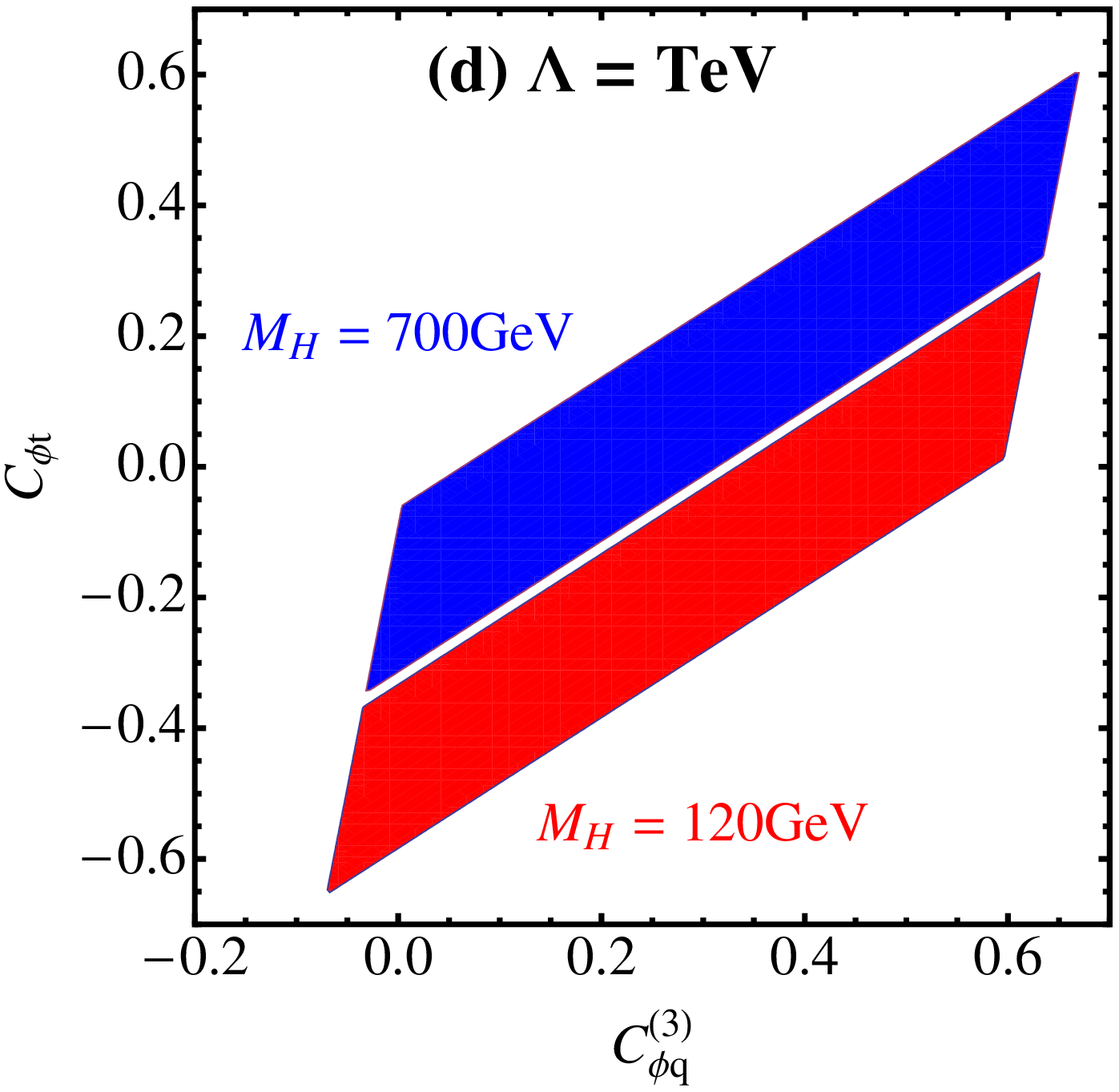}
\includegraphics[clip,scale=0.3]{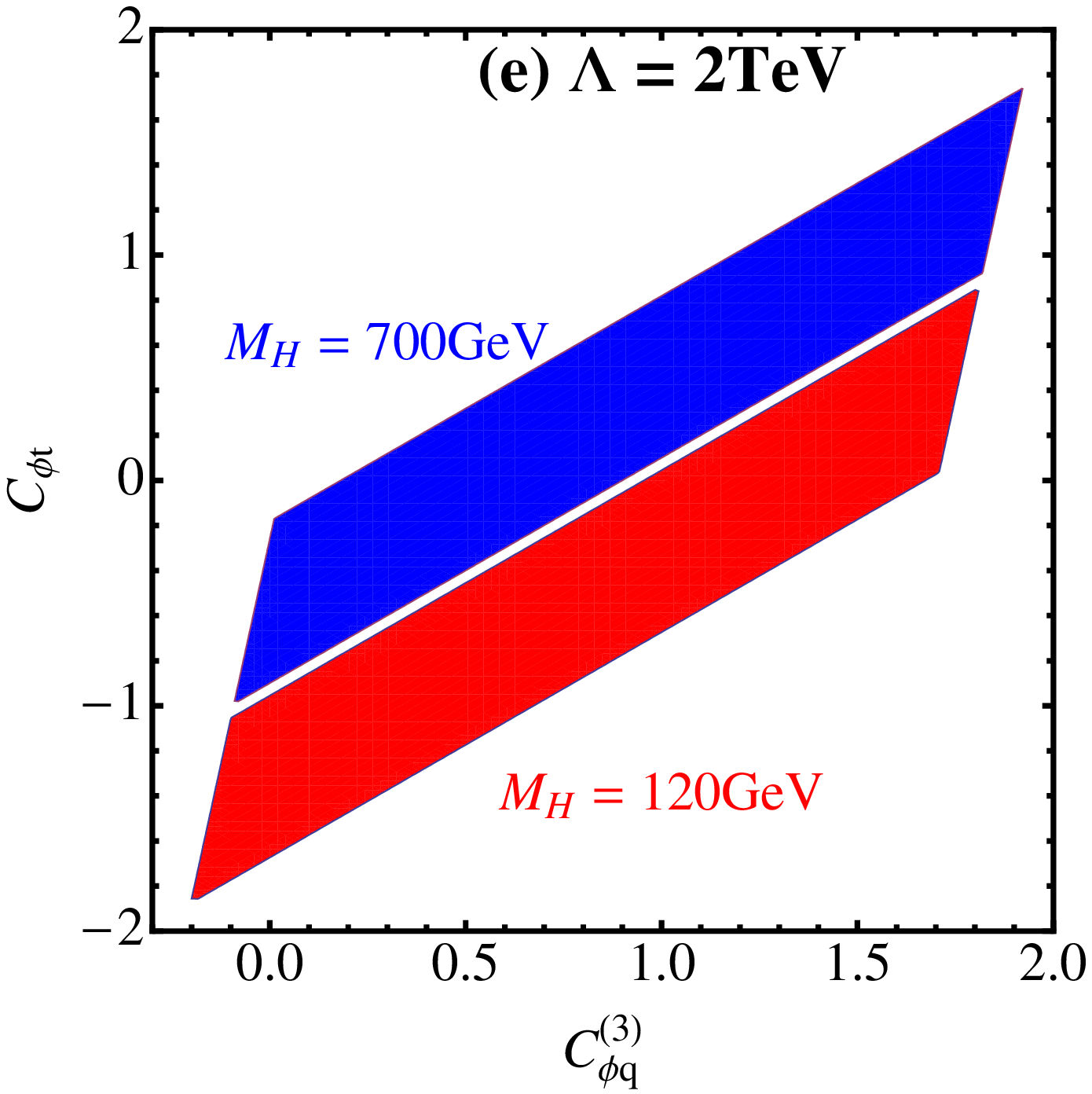}
\includegraphics[clip,scale=0.3]{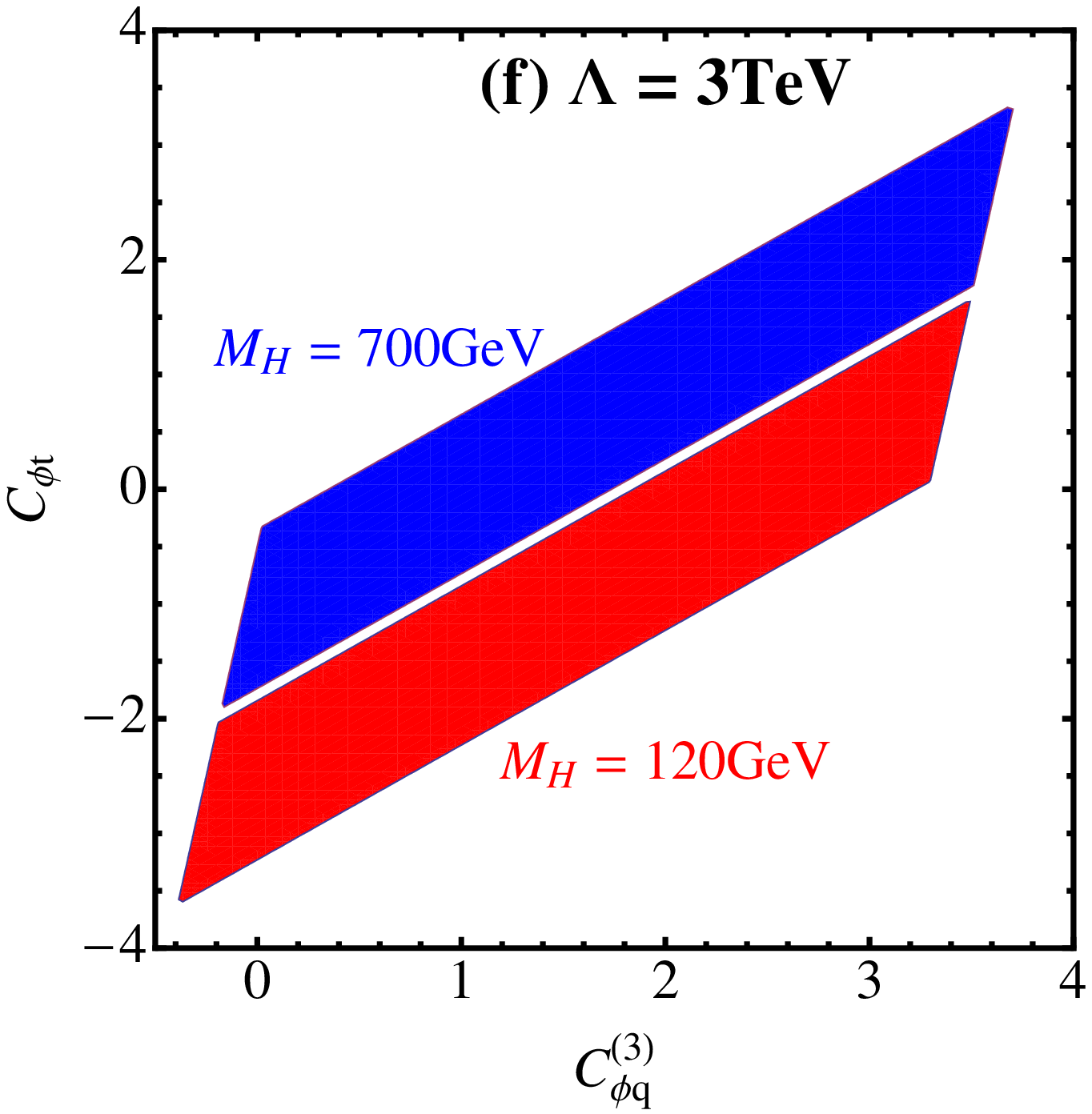}
 
\caption{Allowed regions of $\fl$ and $\fr$ (a-c) and of $c_{\phi q}^{(3)}$
and $c_{\phi t}$ (d-f) for different NP cutoff scales $\Lambda$ and
SM Higgs boson masses $m_{H}$. \label{fig:flfrlimit}}
\end{figure}

Using the experimental results~\cite{Altarelli:2004mr}, \begin{eqnarray*}
4.4\times10^{-3} & \leq\epsilon_{1}^{exp}\leq & 6.4\times10^{-3},\\
-6.2\times10^{-3} & \leq\epsilon_{b}^{exp}\leq & -3.1\times10^{-3},\end{eqnarray*}
we plot the experimental constraints on $\fl$ and $\fr$ in Fig.~\ref{fig:flfrlimit},
assuming the absence of contributions from other operators.
It is important to contrast the allowed region for $\fl$ and $\fr$
with the expected sizes $c_{i}v^{2}/\Lambda^{2}$ based on the power counting
discussed earlier in Section \ref{sect:opana}. As expected, the constraints
from one-loop corrections do not require the magnitudes of $c_{\phi q}^{(3)}$
and $c_{\phi t}$ to be smaller than their ${\cal O}(1)$ natural sizes, see
Fig.~\ref{fig:flfrlimit}(d-f).

Last, we comment on other new physics effects which could modify the above 
bounds. It is worth mentioning that many tree-level induced operators can also
contribute to $Z\to b\bar{b}$ decay, diluting the constraints we derived above
without breaking the correlations between the $\wtb$, $\zbb$ and $\ztt$
effective couplings. For example, four-fermion operators $q\bar{q}b\bar{b}$ 
would affect $Z\to b\bar{b}$ decay via seagull type loop diagrams. 
An exploration of the constraints (or correlations) among all tree-level 
induced operators from the $\rho$ parameter and LEP $Z\to b\bar{b}$ data 
is highly desirable, but it is beyond the scope of this paper and will be
presented elsewhere. Therefore, we take the above constraints as an indication
of the magnitude of the effective couplings but do not limit ourselves to 
the above parameter space in the subsequent collider analysis.

%
%
\section{New Physics Scenarios}

Having established in a model independent fashion the possible pattern of deviations
in the $Wtb$, $Zb{\bar b}$, and $\ztt$ couplings,
we offer a brief survey in this section of new physics models that give rise to such 
deviations. The treatment here is by no means complete and serves only to exemplify
the generality of the operator analysis.

As discussed in the previous section, we are interested mainly in deviations arising from
non-oblique corrections, which are much less constrained experimentally. The simplest
possibilities for such corrections are new particles mixing with the SM top quark, the
SM bottom quark, or both.

The possibility of introducing additional bottom-like quarks ($b^{\prime}$),
which mix with the SM $b$ quark, to 
resolve the discrepancy of the forward-backward asymmetry of
the $b$ quark ($A_{FB}^{b}$) is investigated in Ref.~\cite{Choudhury:2001hs}.
However, 
because of the stringent constraint on $Zb_Lb_L$, the mixing could be significant 
only in the right-handed sector, implying negligible deviations in the $Wt_Lb_L$
coupling. Since the $Wt_Rb_R$ coupling is already severely constrained by 
$b\to s\gamma$ data, as mentioned previously,  
we do not expect this class of models to produce significant
deviations in the $\ztt$ and $Wtb$ couplings.

A custodial $O(3)$ symmetry would protect simultaneously the $\rho$ parameter 
and the $\zbbl$ coupling~\cite{Agashe:2006at}.
In this case significant mixing of $b^\prime$ with the SM $b$ quark in the left-handed
sector is possible, provided additional top-like quarks with appropriate 
quantum numbers are also present so as to cancel the $b'$ contribution to $\zbbl$.
Explicit examples are found in Refs.~\cite{Contino:2006qr,Albrecht:2009xr,Buras:2009ka}.

The third class of models we discuss has an exotic top-like ($t'$) quark which
could mix with the SM top quark. Such a scenario appears quite often in 
theories beyond the SM, especially those attempting to address
the quantum stability of the Higgs boson mass. As is well-known, within the SM the largest
contribution to the one-loop quadratic divergence of the Higgs boson mass comes from
the top quark. Therefore, if the Higgs boson mass is to be at the order of a few hundred GeV
without significant fine-tuning, a new particle (i.e., the $t'$ quark) must be present to cancel
the top quark contribution in the quadratic divergence. This is the case in 
models where the Higgs boson arises as a pseudo-Nambu-Goldstone boson such as in Little
Higgs theories~\cite{ArkaniHamed:2001nc}, the holographic Higgs model~\cite{Contino:2003ve}, 
the twin Higgs model~\cite{Chacko:2005pe}, and so forth. On the other hand, there are also
models with a $t'$ quark which may not deal with the electroweak hierarchy problem explicitly.
Notable examples are those with flat~\cite{ArkaniHamed:1998rs} or 
warped~\cite{Randall:1999ee} extra dimensions where the SM field propagate.
In this case the $t'$ quark is nothing but the Kaluza-Klein partner of the SM top quark, taken to be
the zero mode. Very recently a $t'$ quark was also invoked to explain a possible experimental
excess at the Tevatron ~\cite{Dobrescu:2009vz}.
 
Having surveyed some candidates for new physics beyond the SM, to which our 
model independent analysis applies, we address in the next section how to constrain
the two unknown parameters $\fl$ and $\fr$ at the LHC.

%
%
\section{Measurements of the $Wtb$ and $\ztt$ couplings at the LHC}

In this section we discuss how the $Wtb$ and $\ztt$ couplings could be measured at the LHC.
The former can be measured most directly in single top quark
production and the latter in the $\ztt$ associated production.  We first consider extracting $\fl$
and $\fr$ from total cross section measurements. Then we look beyond the total cross section
and study differential distributions in the decay products of $\ztt$ associated production.

\subsection{Current Experimental Bounds}

Single top quark events result from the $t$-channel process
($ub\to dt$), the $s$-channel process ($u\bar{d}\to t\bar{b}$)
and $Wt$ associated production ($b g \to t W^{-} $). The distinct kinematics
of each of these processes allows differentiation among their contributions.
Observation of single top quark production was reported recently by
the CDF and D0 collaborations~\cite{Abazov:2009ii,Aaltonen:2009jj}.
These results provide the first direct measurement of the product
of the $\wtb$ coupling ($g_{\wtb}$) and the CKM matrix element $V_{tb}$.
CDF quotes $\left|V_{tb}\right|=0.91\pm0.11({\rm stat}+{\rm syst})\pm0.07({\rm theory})$
and a limit $\left|V_{tb}\right|>0.71$ at the $95\%$ C.L. {[}for
$m_{t}=175\,{\rm GeV}${]}. D0 obtains $\left|V_{tb}(1+\fl)\right|=1.07\pm0.12({\rm stat}+{\rm syst}+{\rm theory})$
and a limit of $\left|V_{tb}\right|>0.78$ at the $95\%$ C.L. {[}$m_{t}=170\,{\rm GeV}${]}.
The limit on $V_{tb}$ is derived under assumption of $g_{\wtb}=g_{\wtb}^{SM}$,
i.e. $\fl=0$. However, the bound can also be translated into a bound
on $\fl$ if one assumes the unitarity
of the $3\times3$ CKM matrix element ($\left|V_{tb}\right|=1$).   Taking the D0 result at face 
value and inserting $\left|V_{tb}\right|=1$, we see that these data could allow 
$\fl\simeq {\cal O}(0.1)$, roughly twice the size shown in  Fig.~\ref{fig:flfrlimit}.

On the other hand, obvious limitations
have precluded measurements of the $\ztt$ coupling ($g_{Ztt}$) thus far.
There was insufficient center-of-mass energy at LEP to produce a top
quark pair via $e^{+}e^{-}\to\gamma/Z\to t\bar{t}$. At hadron colliders,
$t\bar{t}$ production is so dominated by the QCD processes $gg\to t\bar{t}$
and $q\bar{q}\to g\to t\bar{t}$ that the signal from $g_{\ztt}$
via $q\bar{q}\to\gamma/Z\to t\bar{t}$ cannot be extracted. However,
one might be able to measure the $g_{\ztt}$ coupling via the process of $gg\to Zt\bar{t}$.

The sensitivity to non-standard
$\wtb$ couplings at the LHC via single top quark production is investigated
in several papers~\cite{Kane:1991bg,Carlson:1994bg,Malkawi:1994tg,Espriu:2001vj,
Espriu:2002wx,Chen:2005vr,Batra:2006iq,Alwall:2006bx}, while a study on extracting
the $\ztt$ coupling in $\ztt$ associated production at the LHC
appears in Refs.~\cite{Baur:2004uw,Baur:2005wi}.
Such measurements would also be a focus of study in a future high energy
electron-positron linear collider (LC). 

It is worth pointing out that, since measurements of single top quark production
can only probe the combination $|V_{tb}(1+\fl)|$, it would be desirable to 
measure $\fl$ independently of $|V_{tb}|$, which could be achieved by
utilizing $\ztt$ associated production.

\subsection{Cross sections for single top quark and $\ztt$ associated production}

Because the operator $\mathcal{O}_{\wtb}$ is proportional to $\fl$,
only the overall normalization of the single top quark cross section
is affected, and, except for normalization, the final state differential distributions
are insensitive. A measurement of $\fl$ requires a very precise measurement
of the total cross section. The coupling $\fl$ also affects top quark
decay, but it does not change the top quark decay branching ratio,
i.e. $Br(t\to W^{+}b)=1$.%
\footnote{Whatever $\fl$ contributes to the matrix element
of the decay is canceled by its modification to the top quark decay width.} 
On the other hand, the $\ztt$ amplitude involves both $\fl$ and $\fr$,
and it is conceivable that differential distributions in the decay products of $\ztt$
production would have different sensitivity to $\fl$ and $\fr$, respectively.
We will look at two possibilities: the opening angle $\Delta\phi(\ell^+,\ell^-)$ from the decay of
$Z\to \ell^+\ell^-$ and the spin correlation between the top quark and the $Z$ decay
products.

The inclusive cross sections for single top quark and $\ztt$ associated
production at the LHC are: 
\begin{eqnarray}
\label{eq:totalcros}
\sigma_{t} & = & \sigma_{t}^{0}\left[1+2\fl+2\dvtb+\mathcal{O}\left(\fl^{2}, \delta V_{tb}^2\right)\right],\label{eq:xsec-singlet-lhc}\\
\sigma_{\ztt} & = & \sigma_{\ztt}^{0}\left[1+4.4\fl-1.5\fr+\mathcal{O}\left(\fl^{2},\,\fr^{2},\,\fl\fr\right)\right],\label{eq:xsec-ztt-lhc}
\end{eqnarray}
where $\sigma_{t}^{0}$ and $\sigma_{\ztt}^{0}$ denote the SM cross
sections for single top quark production and $\ztt$ production,
respectively.   We include the possibility of a non-unitary CKM matrix element 
$\delta V_{tb}=|V_{tb}|^{\rm (exp)} - |V_{tb}|^{\rm (SM)}$. 
From Eq.~(\ref{eq:totalcros}) we see immediately the possibility of extracting 
$\delta V_{tb}$ from 
\begin{equation}
\label{eq:nonvtb}
\dvtb=-0.23 \delta\sigma_{\ztt}+ 0.5 \delta\sigma_{t}-0.34\fr \label{eq:xsec-corr},
\end{equation}
which is not possible from measurements of single top quark production alone.   
In the above, $\fr$ could in principle be measured from differential distributions in 
$\ztt$ associated production, as is discussed in detail below.

Since new physics contributions to the $Wtb$, $\ztt$,  and  
$\zbb$ couplings are of the order $v^2/\Lambda^2 \simeq 1/(16\pi^2)$
for $\Lambda\simeq 1$ TeV, we can safely ignore interference effects between 
new physics and SM one-loop contributions in the total cross section.
Therefore, the SM quantities in Eq.~(\ref{eq:totalcros})
are understood to be evaluated at one-loop level, as calculated
in Refs.~\cite{Sullivan:2002jt,Campbell:2004ch,Cao:2004ky,Sullivan:2004ie,Frixione:2005vw,
Cao:2004ap,Sullivan:2005ar,Campbell:2005bb,Cao:2005pq,Kidonakis:2006bu,Cao:2008af,
Frixione:2008yi,Campbell:2009ss,Lazopoulos:2008de}.

\begin{figure}
\includegraphics[clip,scale=0.5]{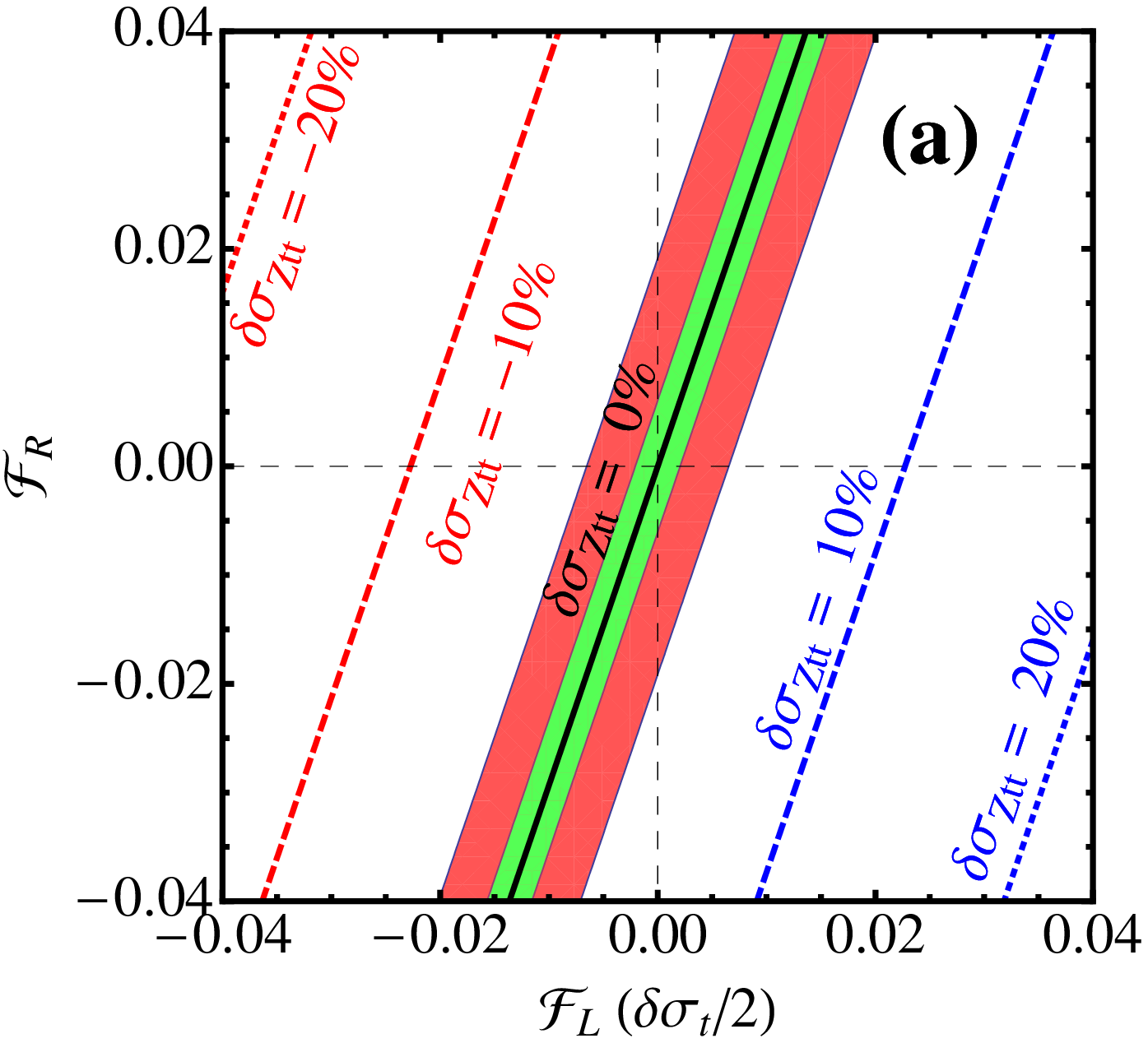}~~%
\includegraphics[clip,scale=0.48]{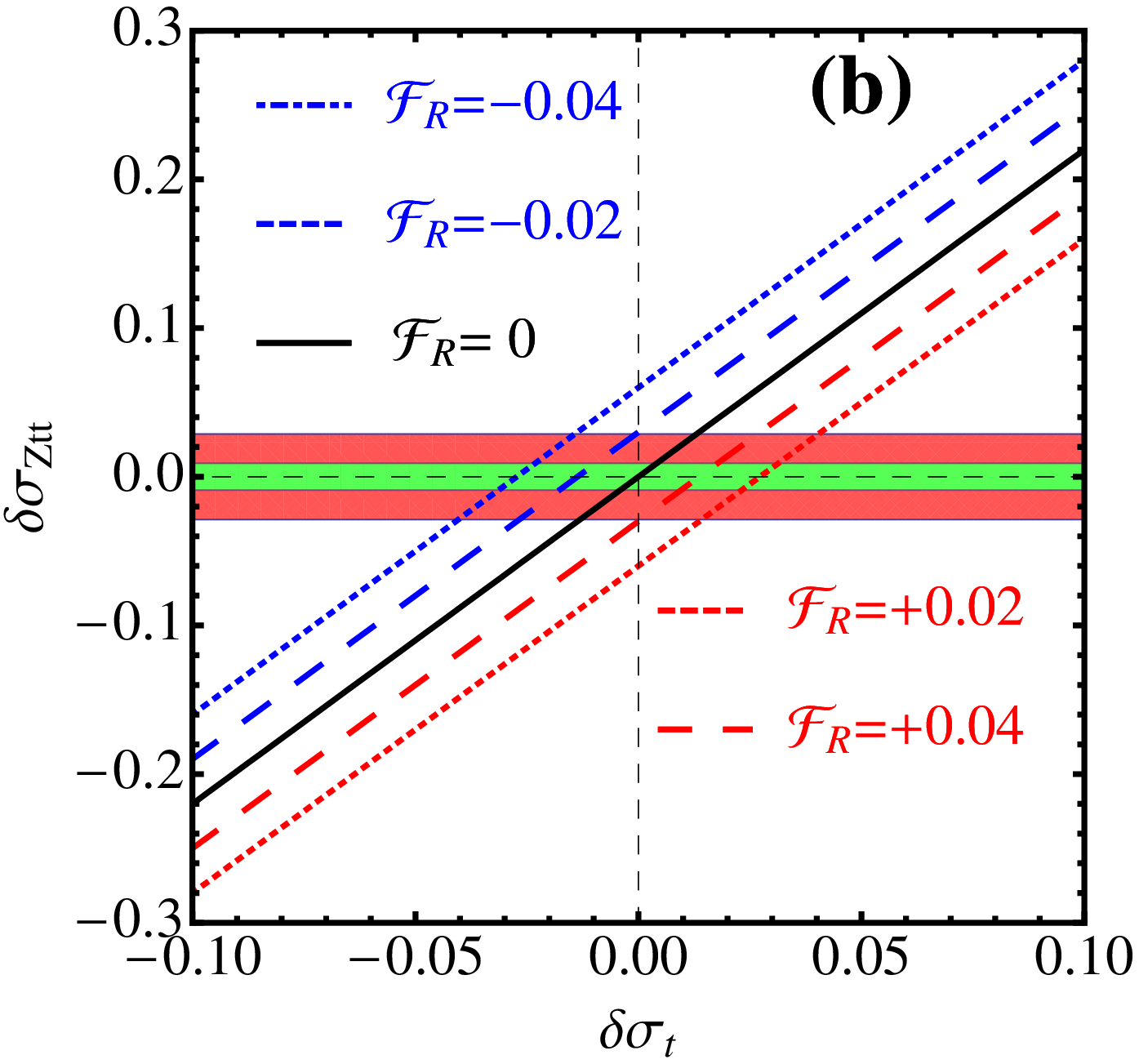}
\caption{(a) Contours of the deviation of the $\ztt$ production cross section
in the plane of $\fl$ and $\fr$, where $\delta\sigma_{\ztt}=-20\%$
(red dotted), $-10\%$ (red dashed), $0\%$ (black), and $10\%$ (blue
dashed); (b) Correlation of $\delta\sigma_{\ztt}$ and $\delta\sigma_{t}$
for different $\fr$. The red (green) band is the 3 standard deviation statistical 
variance of the $\ztt$ production cross section at the LHC with an integrated
luminosity of $10\,{\rm fb}^{-1}$ ($100\,{\rm fb}^{-1}$), respectively. \label{fig:flfr}}
\end{figure}

We define the deviation of the cross sections from the SM predictions
as $\delta\sigma\equiv\left(\sigma-\sigma^{0}\right)/\sigma^{0}.$
The contours of $\delta\sigma_{\ztt}$ in the plane of $\fl$ and
$\fr$ are shown in Fig.~\ref{fig:flfr}(a). The ranges of $\fl$
and $\fr$ in this figure are consistent with the allowed regions shown
in Fig.~\ref{fig:flfrlimit}. All the contours are straight lines
since we keep only the interference term. The anticipated deviation
$\delta\sigma_{\ztt}$ is as large as 20\%. 
We observe that in Eq.~(\ref{eq:xsec-ztt-lhc}) $\fl$ and $\fr$ contribute with opposite signs,
implying partial cancellations in the new physics contributions.
In fact, if $\fl/\fr\simeq1/3$ then $\delta\sigma_{\ztt}\simeq0$, 
as one may see in the black-solid curve of Fig.~\ref{fig:flfr}(a). 
On the other hand, if $\delta\sigma_{\ztt}>0$ we could infer immediately
that $\fl/\fr>1/3$, and vice versa.
The correlation between the deviations of the cross sections
for various values of $\fr$ is revealed in Fig.~\ref{fig:flfr}(b), where $\dvtb=0$ is assumed.
When the right-handed $\ztt$ coupling is not modified by any new physics, say $\fr=0$, then 
\begin{equation}
\delta\sigma_{\ztt}=2.2\delta\sigma_{t},\label{eq:zero_fr}
\end{equation}
as illustrated by the black-solid curve in Fig.~\ref{fig:flfr}(b).
Non-zero values of $\dvtb$ increase the intercept on the $y$-axis of all the lines
in Fig.~\ref{fig:flfr}, as $\dvtb$ is always negative.
The solid bands in the figure are the 3 standard deviation statistical variations of the SM total 
cross section for different luminosities.   
Given large enough deviations in both $\ztt$ and single top quark production, 
one could determine the values of $\fl$ and $\fr$ uniquely if $V_{tb}=1$. 
(For example, one can determine $\fr$ by substituting $\fl$ derived from the single top quark
measurement into the $\ztt$ measurement.) 
If one relaxes the constraint on $V_{tb}$ then
the two measurements merely yield a relation between 
$V_{tb}$ and $\fr$, see Eq.~(\ref{eq:nonvtb}).
Nevertheless, it may be possible to extract $\fr$ independently from   
differential distributions in $\ztt$, as these are sensitive to the anomalous couplings.  
If one takes $\delta V_{tb}=0$ as an assumption, then
$\fl$ and $\fr$ could be over-constrained by measurements of the total and
differential cross sections, allowing for a check on the relation in 
Eq.~(\ref{eq:CC-NC}).   
In the rest of this study we simply assume
$\delta V_{tb}=0$ unless otherwise specified.

Next we consider two differential distributions in
$\ztt$ associated production: the opening angle $\Delta\phi(\ell^+\ell^-)$ in 
$Z\to \ell^+\ell^-$ and spin correlations defined below.  
In studying the impact of the two effective couplings $\fl$
and $\fr$ on the top quark spin correlations, we look at the final state  
\begin{equation}
q\bar{q}/gg\to t\bar{t}Z,\, 
Z\to\ell^{\prime+}\ell^{\prime-},\, 
t\to bW^{+}(\to\ell^{+}\nu),\,
\bar{t}\to\bar{b}W^{-}(\to jj),
\label{eq:signal}\end{equation}
where $\ell(\ell^{\prime})$ denotes a charged lepton, $b$ a bottom
quark jet, and $j$ a light quark jet. The collider signature is 
$\ell^{\prime+}\ell^{\prime-}\ell^{+}b\bar{b}jj$
plus missing energy $\met$. The backgrounds to the tri-lepton final
state considered in Ref.~\cite{Baur:2004uw} are from $(t\bar{b}Z+\bar{t}bZ)+X$
production, e.g. $(t\bar{b}Z+\bar{t}bZ)jj$ and $(t\bar{b}Z+\bar{t}bZ)\ell\nu$,
and from non-resonant $WZb\bar{b}jj$ production. As evaluated in
\cite{Baur:2004uw}, both background rates are one order of magnitude
less than $Zt\bar{t}$ production, and we do not consider them in
this study.  Heavy flavor contributions to tri-lepton final states are 
examined in Ref.~\cite{Sullivan:2008ki}.

We begin with the initial expectation that the spin of the top quark
in $\ztt$ production is a good discriminating variable. 
The potential for measuring the top quark polarization
in $\ztt$ production is not studied yet in the literature, to the best of our knowledge.
The charged lepton in top quark decay can be used to measure the top quark spin
and thereby determine the anomalous couplings. 
To investigate this possibility
quantitatively, we perform a quantitative numerical simulation of
$\ztt$ associated production including the decays of the $Z$ boson
and top quarks. 
In addition to the usual helicity basis, we propose and examine a new 
`optimal' basis that improves the measurement of top quark polarization.

Another differential observable is the opening angle
in the transverse plane between the two charged leptons from $Z$
boson decay, $\Delta\phi(\ell^{\prime+}\ell^{\prime-})$, also investigated
in the study of Refs.~\cite{Baur:2004uw,Baur:2005wi}.
In this work we show explicitly that this azimuthal opening angle
is a better observable for measuring the anomalous $\ztt$ couplings than the 
top quark spin because it is relatively insensitive to kinematic cuts. 
The correlation between $\ztt$ and single top quark production is exploited 
at the end of this section.

We mention in passing that $\ztt$ associated
production is an important SM background in searches for possible
new physics, especially when the $Z$ boson decays into neutrinos,
resulting in missing transverse energy in collider detectors.
Examples are 
$pp\to\tilde{t}\tilde{t}\to t\bar{t}\tilde{\chi}\tilde{\chi}$ in the minimal 
supersymmetric standard model (MSSM) where $\tilde{t}$ and $\tilde{\chi}$ 
denote the top squark and the neutralino (Wino like), and 
$pp\to T_{-}\bar{T}_{-}\to t\bar{t}A_{H}A_{H}$ in the Little Higgs theories with
T-parity (LHT)~\cite{Cheng:2003ju,Low:2004xc,Cheng:2004yc,Hubisz:2004ft,Hubisz:2005tx,Belyaev:2006jh},
where $T_{-}$ and $A_{H}$ are the T-odd top quark and photon partners,
respectively. In the LHT, the $A_{H}-t-T_{-}$ coupling is predominately
right-handed polarized. In the MSSM the $\tilde{\chi}-t-\tilde{t}$ coupling 
depends on the $\tilde{t}$ mixing. The top quark polarization
then will be a key to distinguish or to provide information on various models.

%
%
\subsection{Monte Carlo Simulation}

We perform a Monte Carlo simulation of $Zt{\bar{t}}$
production at the parton level, sufficient for our purposes. 
We do not include SM one-loop contributions in our simulation of the differential 
decay rate.   These effects should certainly
be taken into account when one attempts to analyze real data.

\subsubsection{Event Reconstruction}

To mimic detector capability, we require the transverse momentum of the charged
lepton and jets (including both $b$ and $j$) to satisfy the following
basic cuts: 
\begin{eqnarray}
 &  & p_{T}^{\ell}>15\,{\rm GeV},\quad\left|\eta_{\ell}\right|<2.5,
	  \quad p_{T}^{b}>20\,{\rm GeV},\quad\left|\eta_{b}\right|<2.5,\nonumber \\
 &  & p_{T}^{j}>15\,{\rm GeV},
	  \quad\left|\eta_{j}\right|<2.5,\quad\met>20\,{\rm GeV},\nonumber \\
 &  & \Delta R(j,\, j)>0.4,\quad\Delta R(j,\,\ell)>0.4,
	  \quad\Delta R(j,\, b)>0.4,\quad\Delta R(b,\, b)>0.4.\label{eq:cut}
\end{eqnarray}
Here $\Delta R=\sqrt{(\Delta\phi)^{2}+(\Delta\eta)^{2}}$ is the
separation in pseudo-rapidity-azimuth space, and $\met$ is the missing
transverse momentum originating from the neutrino which escapes the
detector. In this study we adopt the $p_{T}$-dependent $b$-tagging
efficiency defined as \cite{Carena:2000yx}
\[ \epsilon_{b}=0.57\times\tanh\left(\frac{p_{T}^{b}}{35\,{\rm GeV}}\right).\]
We smear all final state parton momenta by a Gaussian distribution with 
\[ \frac{\Delta E}{E}=\frac{50\%}{\sqrt{E}},\] 
where $E$ is the energy of the observed parton, and the resolution
in energy is assumed to be $50\%\sqrt{E}$.
We also require that there be a same flavor, opposite-sign charged lepton
pair with invariant mass near the $Z$ resonance, 
\[ \left|m_{\ell^{\prime+}\ell^{\prime-}}-m_{Z}\right|<10\,{\rm GeV}.\]
As a result of this final state signature requirement, $\ztt$ production
as observed is insensitive to $t\bar{t}\gamma$ production, where
$\gamma$ denotes a virtual photon.

To study spin correlations, one must reconstruct the $W$-boson pair
and the top quark pair. The hadronically decaying $W$ could be reconstructed 
from the invariant mass of the two light jets, while the leptonically decaying
$W$-boson is reconstructed from the final
state electron and the observed missing transverse energy $\met$.
The lack of information about the longitudinal component of the neutrino
momentum ($p_{z}^{\nu}$) is addressed by requiring the invariant
mass of the electron-neutrino system to be equal to the mass of the
$W$-boson. This additional constraint yields two possible solutions
for $p_{z}^{\nu}$, and typically, both of them are physical solutions
for a signal event. We follow the prescription in Ref.~\cite{Kane:1989vv}
to choose the solution which has the smaller $\left|p_{Z}^{\nu}\right|$.
This method picks the correct $p_{z}^{\nu}$ in about $70\%$ of the
events passing the above basic cuts. We find no physical solution
for quite a few events due to the detector smearing effects. To recover
these events, we generate a Breit-Wagner distribution around $m_{W}$
and use the generated mass to derive $p_{Z}^{\nu}$. About $7\%$
of the remaining events do not exhibit a physical solution and are
not included in our analysis. 

\begin{figure}
\includegraphics[scale=0.3]{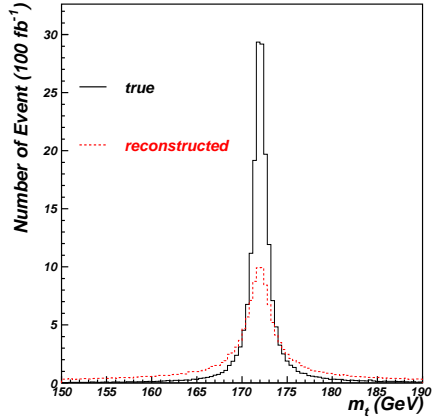}
\caption{
Invariant mass distribution of the true top quark (black) and the reconstructed top (red). 
An integrated luminosity of $100\,{\rm fb}^{-1}$ is used here. 
\label{fig:w-top-recst}}
\end{figure}

To reconstruct the top quark, we combine the reconstructed $W$-boson
with the $b$-jet from the top quark decay. The challenge in this
case is to identify the correct jet. To this end we make use of the top quark
mass measured in $t\bar{t}$ events.
In single top quark events,
the $Wj$ combination that gives an invariant mass closest to the
true top mass is chosen as the reconstructed top quark. In $\ztt$
events there are two $W$ bosons and two $b$-jets in the final state.
Labeling the leptonically decaying $W$ boson $W_{\ell}$ and the
hadronically decaying $W$-boson $W_{h}$, we loop over the
combinations of the $W$-bosons and $b$-jets, i.e. ($W_{\ell}b_{1},\, W_{h}b_{2}$)
and ($W_{\ell}b_{2},\, W_{\ell}b_{1}$), and we calculate the invariant
masses of the reconstructed top quarks. We then require all the masses
of the reconstructed $(Wb)$ systems to be within $20\,{\rm GeV}$
of the top quark mass $m_{t}=173.1\,{\rm GeV}$~\cite{tevatron:2009ec}, i.e.
\[
	\left|m_{Wb}-m_{t}\right|<20\,{\rm GeV}.
\]
We calculate the deviations from the true top quark mass ($m_{t}$)
for each combination, 
\[
	\Delta_{ij}=\sqrt{(m_{W_{\ell}b_{i}}-m_{t})^{2}+(m_{W_{h}b_{j}}-m_{t})^{2}},
\]
and select the combinations with the minimal deviations to be the
reconstructed top quark pair. This simple algorithm works well. It
picks the correct combination about $99\%$ of the time.

Our analysis is limited mainly by the neutrino reconstruction and
experimental uncertainties. For example, due to neutrino reconstruction,
the top quark reconstructed from the $be^{+}\met$ system exhibits
a much broader mass spectrum compared to the anti-top quark reconstructed
from three jet system of $bjj$, as seen in Fig.~\ref{fig:w-top-recst}.

\subsubsection{$\Delta\phi(\ell^+\ell^-)$ and Spin Correlations}

\begin{figure}
\includegraphics[scale=0.5]{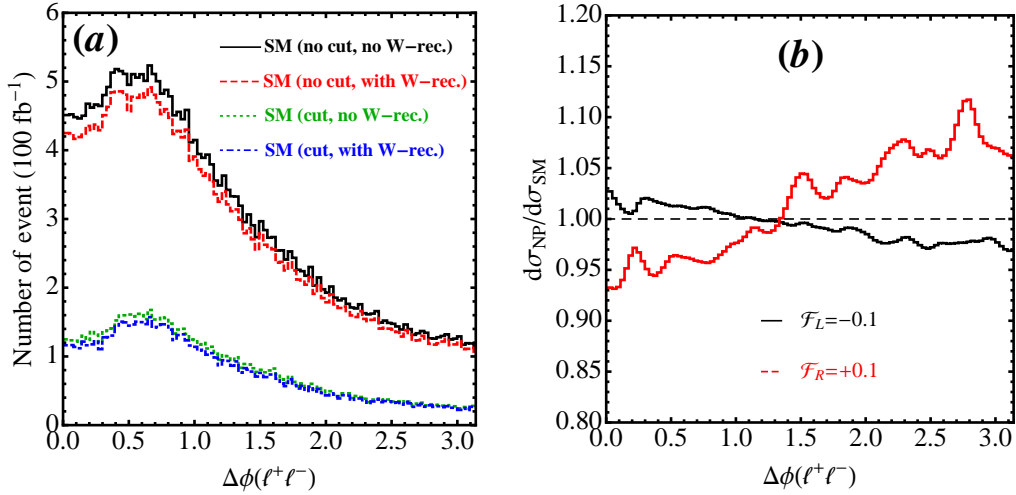}
\caption{(a) The SM distribution of $\Delta\phi_{\ell\ell}$, where the black
curve denotes the distribution at the partonic level without any kinematic
cuts, the red curve labels the distribution after neutrino reconstruction
without kinematic cuts, the green curve represents the partonic
distribution after kinematic cuts, and the blue curve shows the distribution
after kinematic cuts and neutrino reconstruction; 
(b) Ratio of the new physics cross section to the SM result in the $\Delta\phi_{\ell\ell}$ 
distribution, $\fl=-0.1$ and $\fr=0$ (black), and $\fr=0.1$ and $\fl=0$ (red).   
The fluctuations are caused by statistics.
 \label{fig:dphidist}}
\end{figure}

An observable is needed which changes shape in the presence of the
anomalous couplings $\fl$ and $\fr$. The opening angle between the
two charged leptons from the $Z$ boson decay $Z\to\ell^{\prime+}\ell^{\prime-}$
in the transverse plane, $\Delta\phi(\ell^{\prime+}\ell^{\prime-})$,
is such a candidate as is pointed out in Ref.~\cite{Baur:2004uw}.
Since $\Delta\phi_{\ell\ell}$ pertains to the $Z$ boson, it does
not depend on the neutrino reconstruction. To illustrate this point,
we plot the $\Delta\phi_{\ell\ell}$ distribution in the SM in Fig.~\ref{fig:dphidist}(a).
Neutrino reconstruction reduces the number of observed events but
does not change the shape of distribution. The kinematic cuts suppress
the number of events but also do not change the shape.

The sensitivity of $\Delta\phi_{\ell\ell}$ to the anomalous couplings
is shown in Fig.~\ref{fig:dphidist}(b) where we choose the somewhat
generous values $\fl=-0.1$ and $\fr=0.1$ for illustration. Note that 
values of this magnitude are not inconsistent with the Fermilab collider 
data on single top quark production, mentioned at the beginning of this 
section, although for such large $\fl$ and $\fr$ there must be additional contributions, 
other than those from the operators considered in Section II,
in the precision electroweak measurements so as to relax the bounds
in Fig.~\ref{fig:flfrlimit}.

\begin{figure}
\includegraphics[scale=0.5]{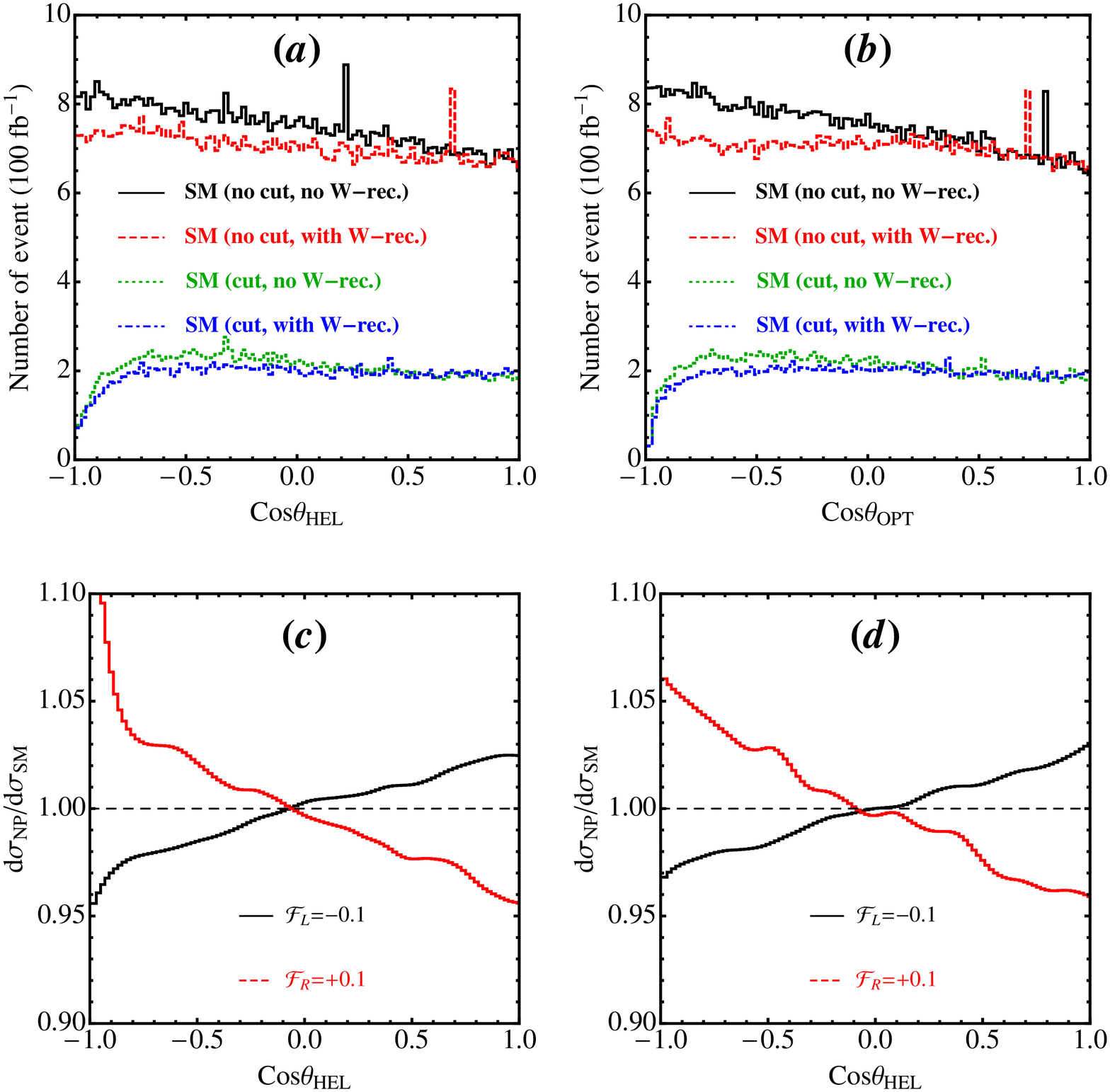}
\caption{The upper panels show the SM distribution of $\cos\theta_{hel}$ (left)
and $\cos\theta_{opt}$ (right) where the meaning of colored curves
is the same as Fig.~\ref{fig:dphidist}. 
The lower panels show the
ratio of the new physics cross section to the SM value in the distributions of 
$\cos\theta_{hel}$ (left) and $\cos\theta_{opt}$(right) after kinematic cuts and
neutrino reconstruction: black for $\fl=-0.1$ and $\fr=0$, and red for $\fr=0.1$
and $\fl=0$. Here, $\theta_{hel}$($\theta_{opt}$) is the angle between the positron
and the top quark spin in the top quark rest frame in the helicity (optimal) basis,
respectively.
\label{fig:cthdist}}
\end{figure}

The top quark spin asymmetry is another observable sensitive to the
anomalous couplings. At the LHC $\ztt$ production proceeds predominately
through the gluon fusion process $gg\to\ztt$.  The presence of left- and 
right-handed couplings of the $Z$ to the top quark in $\ztt$ production 
means that parity is slightly broken, resulting in a small top quark spin
asymmetry.  The anomalous couplings $\fl$
and $\fr$ might amplify or weaken the parity violation effects, and
a measurement of the asymmetry might provide a good probe for these
anomalous couplings. Among the decay products of the top quark, the
charged lepton is maximally correlated with the top quark spin~\cite{Mahlon:1995zn,Parke:1996pr}.
We can thus obtain the most distinctive distribution by plotting the
angle between the spin axis and the charged lepton in the reconstructed top quark
rest frame. Different choices of the reference frame to define the
top quark polarization are found in the literature. In the helicity
basis the top quark spin is measured along the top quark direction
of motion in the center of mass (c.m.) frame. However, one must bear
in mind that the anomalous couplings affect the production of the
$Z$ boson and top quark pair but not the top quark decay. Therefore,
these couplings can be probed better in the spin correlation between
the top quark decay products and the $Z$ boson decay products. We
find that the choice of the negatively charged lepton from the $Z$
boson decay to measure the top quark spin direction amplifies the
top quark spin correlation effects by almost a factor of 2.

In Fig.~\ref{fig:cthdist} we show the $\cos\theta$ distribution
for different spin bases, where $\cos\theta$ is defined as 
\begin{equation}
\cos\theta=\frac{\vec{s}_{t}\cdot\vec{p}_{\ell}^{*}}{|\vec{s}_{t}||\vec{p}_{\ell}^{*}|}.
\label{eq:cth}
\end{equation}
Here $\vec{s}_{t}$ is the three-momentum of the top quark spin in
the reconstructed $\ztt$ c.m. frame, and $\vec{p}_{\ell}^{*}$
is the charged lepton three-momentum defined in the rest frame of
the top quark. In the helicity basis $\vec{s}_{t}$ is chosen to be
the direction of the top quark in the c.m. frame, while in the {}``optimal''
basis $\vec{s}_{t}$ is along the momentum of the negatively charged lepton
from the $Z$ boson decay in the top quark rest frame.

The angular distributions in Fig.~\ref{fig:cthdist}(a,b) show a
clear slope before $W$ reconstruction and kinematic cuts are imposed.
However, the expected top quark spin correlations are diluted by the
reconstruction of the neutrino momentum $p_{Z}(\nu)$ and the kinematic
cuts, as shown in Fig.~\ref{fig:cthdist}.   Affected significantly by kinematic 
cuts, the top quark spin asymmetry seems not the best
choice of variable for measuring the anomalous couplings. It may be
of relevance only once LHC experiments reach a very high level of
accuracy.

%
%

\subsection{Projected Bounds}

\subsubsection{LHC Reach}

We use the results of the event simulation outlined in Section IV B to 
derive projected bounds on deviations from the SM.  The bounds are 
obtained here primarily from our fits to the distribution in $\Delta\phi_{\ell\ell}$.  
Our simulation is done at the leading order level. 
Theoretical uncertainties, arising from the uncalculated higher order corrections
and from the parton distribution functions, should be included in order to make a fully 
realistic prediction.   Next-to-leading 
order (NLO) QCD corrections reduce the renormalization/factorization scale
dependence to $\sim 10\%$ after the choice of an appropriate scale~\cite{Lazopoulos:2008de}, 
and they potentially affect final state differential distributions.  
While a detailed simulation at NLO is in order, it is beyond the scope of this work 
and we leave it to future work.
Here, we conservatively consider an uncertainty of $30\%$ on our $\ztt$ production 
estimates.   The bounds will be improved when a more accurate NLO simulation is 
available.

\begin{figure}
\includegraphics[scale=0.5]{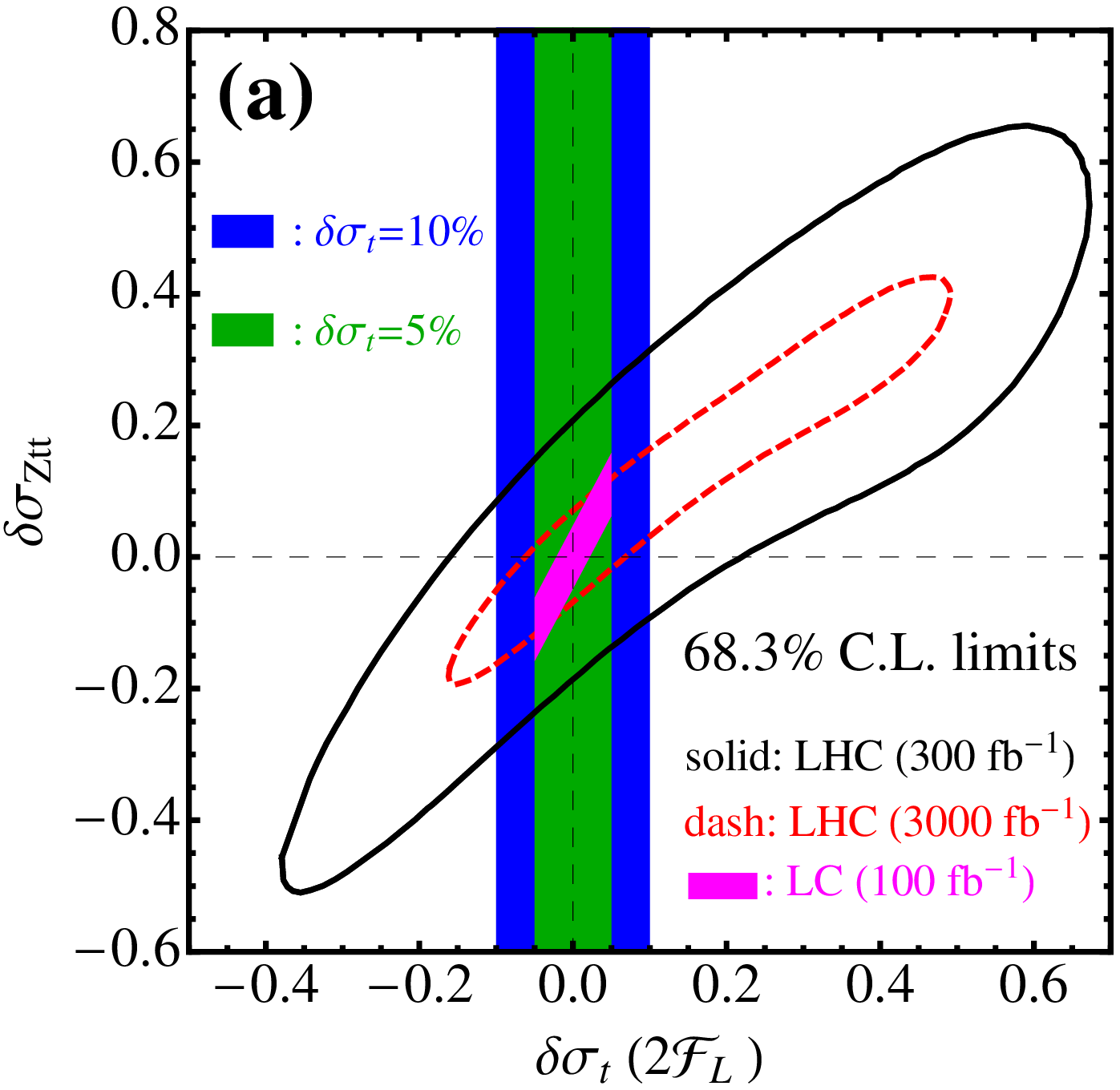}~~~~%
\includegraphics[scale=0.5]{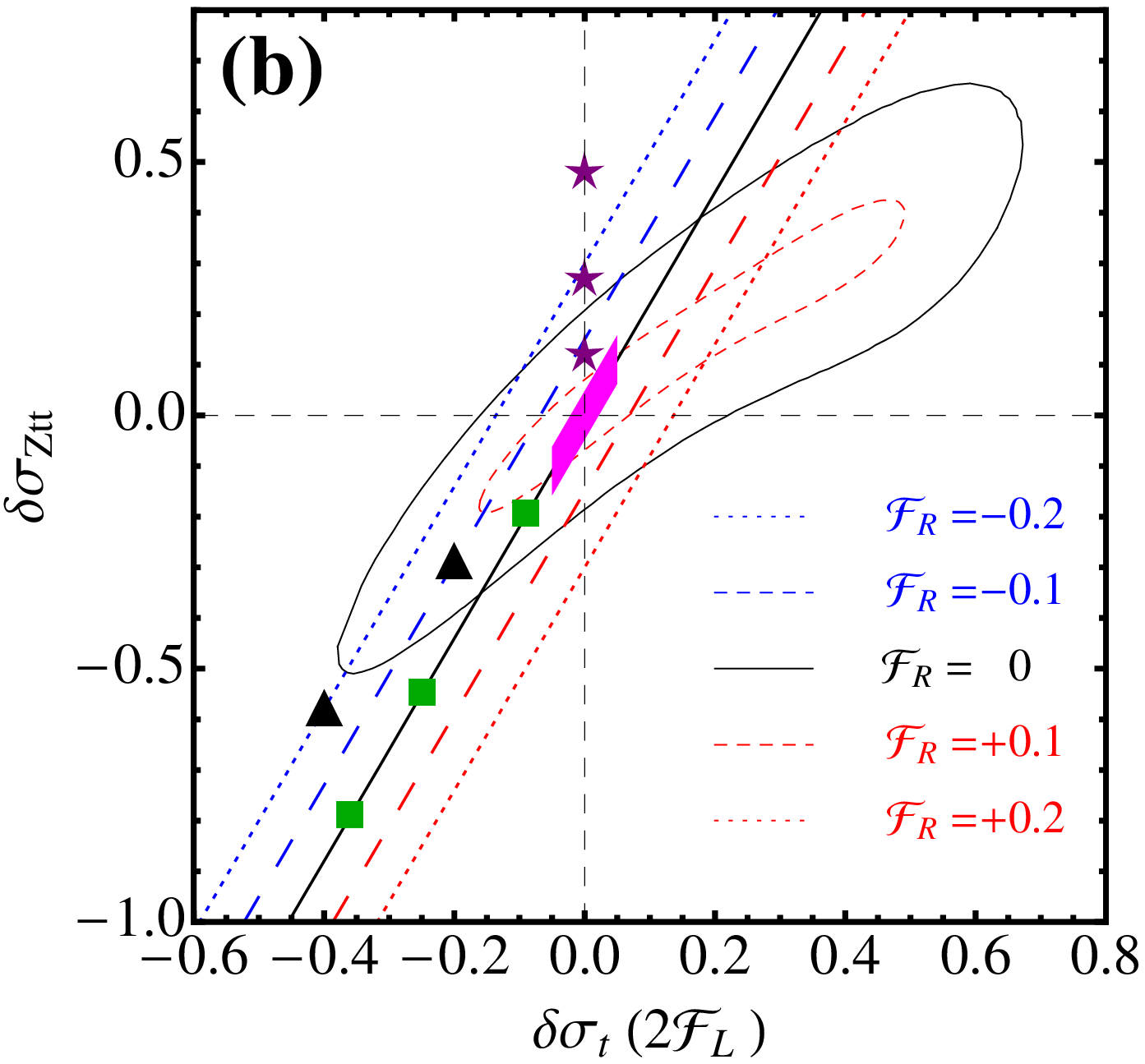}
\caption{(a): Projected $68.3\%$ C.L. bounds on the deviation of the cross
section for $\ztt$ production at the LHC with an integrated
luminosity of $300\,{\rm fb}^{-1}$(solid) and $3000\,{\rm fb}^{-1}$(dashed)
and from a LC ($\sqrt{s}=500\,{\rm GeV}$) with an integrated luminosity
of $100\,{\rm fb}^{-1}$ (magenta band). The expected accuracy of the 
single top quark cross section is represented by
the vertical bands, green (blue) for $\delta\sigma_{t}<\pm5(\pm10)\%$).
(b): Various new physics model predictions, where the box (star, triangle) denotes
the left-handed $t^{\prime}$ (right-handed $t^{\prime}$, sequential
fourth generation) model, respectively. See text for details. \label{fig:lhcreach}}
\end{figure}

In Fig.~\ref{fig:lhcreach} we display the deviation of the $\ztt$ cross section 
along one axis and the deviation of the single top quark production cross 
section along the other.  Figure~\ref{fig:lhcreach} (a) shows the projected 
$68\%$ C.L. bounds on the $\ztt$ production cross section for integrated 
luminosities of $300\,{\rm fb}^{-1}$ (solid) and $3000\,{\rm fb}^{-1}$ (dashed) 
at the LHC. The vertical bands denote the deviation of the single top quark production 
cross section: green for $\left|\delta\sigma_{t}\right|\leq5\%$ and blue
for $\left|\delta\sigma_{t}\right|\leq10\%$.   The straight lines in 
Fig.~\ref{fig:lhcreach}(b) demonstrate the strong correlation between 
$\ztt$ production and single top quark production induced by vanishing deviation 
of  the $\zbb$ coupling.   The black solid line denotes $\fr=0$, see 
Eq.~(\ref{eq:zero_fr}).   Non-zero $\fr$ will shift the curve up ($\fr<0$) or 
down ($\fr>0$), see blue (red) lines. 

The expectations of various new physics models are also plotted in Fig.~\ref{fig:lhcreach}(b). 
The top prime model~\cite{Dobrescu:2009vz} predictions are labeled by the  
(green) squares, where the mixing angle $s_{L}=0.3,\,0.5,\,0.6$ from top to bottom
corresponding to $\fl=-0.045,\,-0.125,\,-0.18$, respectively.  We
also consider a right-handed $t^{\prime}$ model~\cite{Chivukula:1998wd} in which only 
the right-handed $\ztt$ coupling is modified.   Its expectations are shown 
as the (purple) stars, where the mixing angle $s_{R}=0.8,\,0.6,\,0.4$ from top to bottom
corresponding to $\fr=-0.32,\,-0.18,\,-0.08$, respectively.   Another model includes 
a sequential fourth generation~\cite{Kribs:2007nz} whose quarks
mix substantially with the third family.  Both the $\zttl$ and the $\zttr$
couplings are modified by mass mixing of the top quark and fourth
generation up-type quark ($u^{4}$).   One has to assume no
mixing of the bottom quark in order to protect the $\zbb$ coupling.
When $u_{L}^{4}$ and $u_{R}^{4}$ are degenerate as it would be true 
in our effective operator framework, $\fl=\fr$,  shown as the (black) triangles,
$s_{L}=s_{R}=0.44,\,0.63$ ($\fl=\fr=-0.1,\,-0.2$) from top to bottom. 

Figure~\ref{fig:lhcreach} is made under the assumption that the relation in Eq.~(\ref{eq:CC-NC}) is valid.  
If instead one is interested checking Eq.~(\ref{eq:CC-NC}) as a prediction of
$SU(2)_L\times U(1)_Y$ symmetry,
one has to measure $\fl$ and $\fr$ to a high precision from two independent measurements,
in order to verify $\mathcal{F}_{L}^{(\ztt)}=2\mathcal{F}_{L}^{(Wtb)}$.
Such measurements may have to await a LC where one can measure the 
top quark polarization
by choosing the polarization of the incoming electron beam. 
Since the SM gauge symmetry is well-established so far,
we simply invoke Eq.~(\ref{eq:CC-NC}) to analyze data at the LHC.  One
can still gain important information at the LHC:
\begin{itemize}

\item Assuming $\delta V_{tb}=0$, one 
can use the single top quark production measurement to determine $\fl$.  Then
$\fr$ can be determined from the $\ztt$ total cross-section measurement uniquely. 

\item If $\fl$ and $\fr$ are measured from the differential distributions in $\ztt$ production,
one could then disentangle $|V_{tb}|$ from $\fl$ in single top quark production. This result 
could allow us to determine whether $\delta V_{tb}\neq 0$.

\item The sign of the anomalous couplings as determined by the LHC data could carry
important information in terms of distinguishing different classes of models.
For example, if the anomalous couplings are induced by the mixing of SM
particles with heavy exotic particles, then the sign is negative (relative
to the SM coupling) due to the mixing matrix.  Observation
of a positive anomalous coupling 
would imply that either the new physics model is a  strongly interacting theory or the 
third generation quarks are in a
higher representation of the SM gauge group $SU(2)_{L}\times U(1)_{Y}$~\cite{Agashe:2006at}.
\end{itemize}
It is convenient to summarize the
above results in the plane of anomalous $\ztt$ couplings $\fl$ and
$\fr$ as shown in Fig.~\ref{fig:LCreach}. The new physics models
discussed above are distributed in different regions in the plot. 

\subsubsection{Linear Collider Reach}

Here we comment briefly on the reach in a linear collider.
The anomalous $\ztt$ coupling
could be measured in $e^{+}e^{-}\to\gamma/Z\to t\bar{t}\to b\ell^{+}\nu\bar{b}jj$
at a high energy electron-positron linear collider.   Our LHC results in
Fig.~\ref{fig:lhcreach}(a) may be compared with expectations presented
in the American Linear Collider working group report~\cite{Abe:2001nqa}
for a 500 GeV machine and 100 fb$^{-1}$ luminosity. The electron beam
is assumed to be $80\%$ polarized. It is estimated that $\fl$ and
$\fr$ can be measured to $3.7\%$ and $3.2\%$ accuracy in $\ztt$
production, and that $\fl$ can be measured to $2.5\%$ in single
top quark production. The magenta band in Fig.~\ref{fig:lhcreach}(a)
denotes the 1 standard deviation bound on the $\ztt$ production cross section at
a LC. Furthermore, a study of top quark anomalous coupling measurements
via single top quark production at an electron-photon collider
shows that $\fl$ can be measured to $1\%$ with 500 fb$^{-1}$
in a TeV machine~\cite{Cao:2006pu,Sahin:2007tz}.   The figures suggest 
that the LHC with $\mathcal{L}=3000\,{\rm fb}^{-1}$ can reach a similar
accuracy in $\ztt$ production, together with precise measurement of 
the single top quark production cross section, as a 
LC ($\sqrt{s}=500\,{\rm GeV}$ and $\mathcal{L}=100\,{\rm fb}^{-1}$).  

As mentioned previously, such high precision at a linear collider could
allow us to test the relation in Eq.~(\ref{eq:CC-NC}) as a prediction of the
electroweak symmetry, if one examines sufficiently many differential observables.
In addition, the combination of data from the LHC and a LC may also
allow for a precise determination of $|V_{tb}|$, which would not be possible using
single top quark production alone.

\begin{figure}
\includegraphics[scale=0.5]{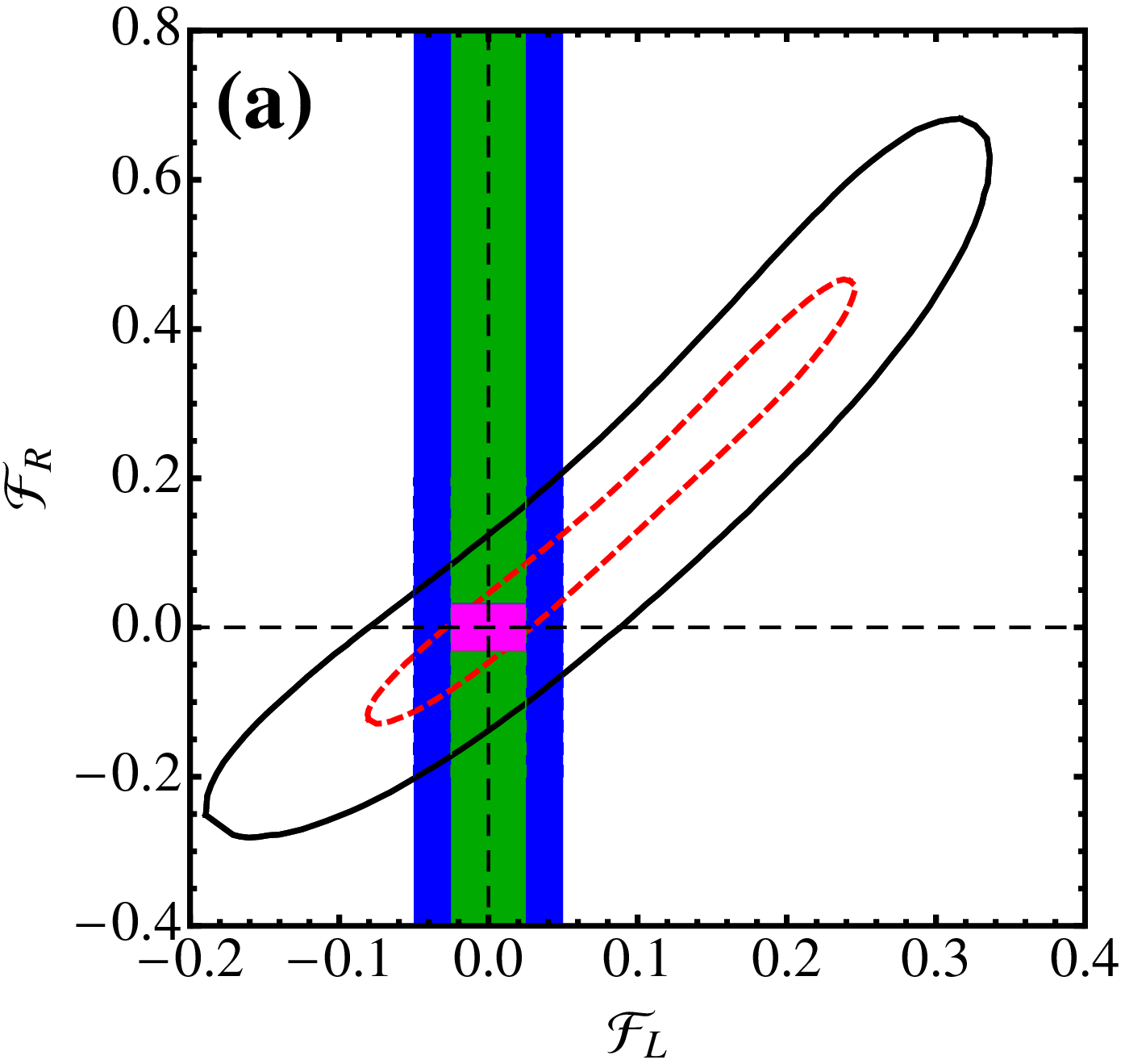}~~~~%
\includegraphics[scale=0.5]{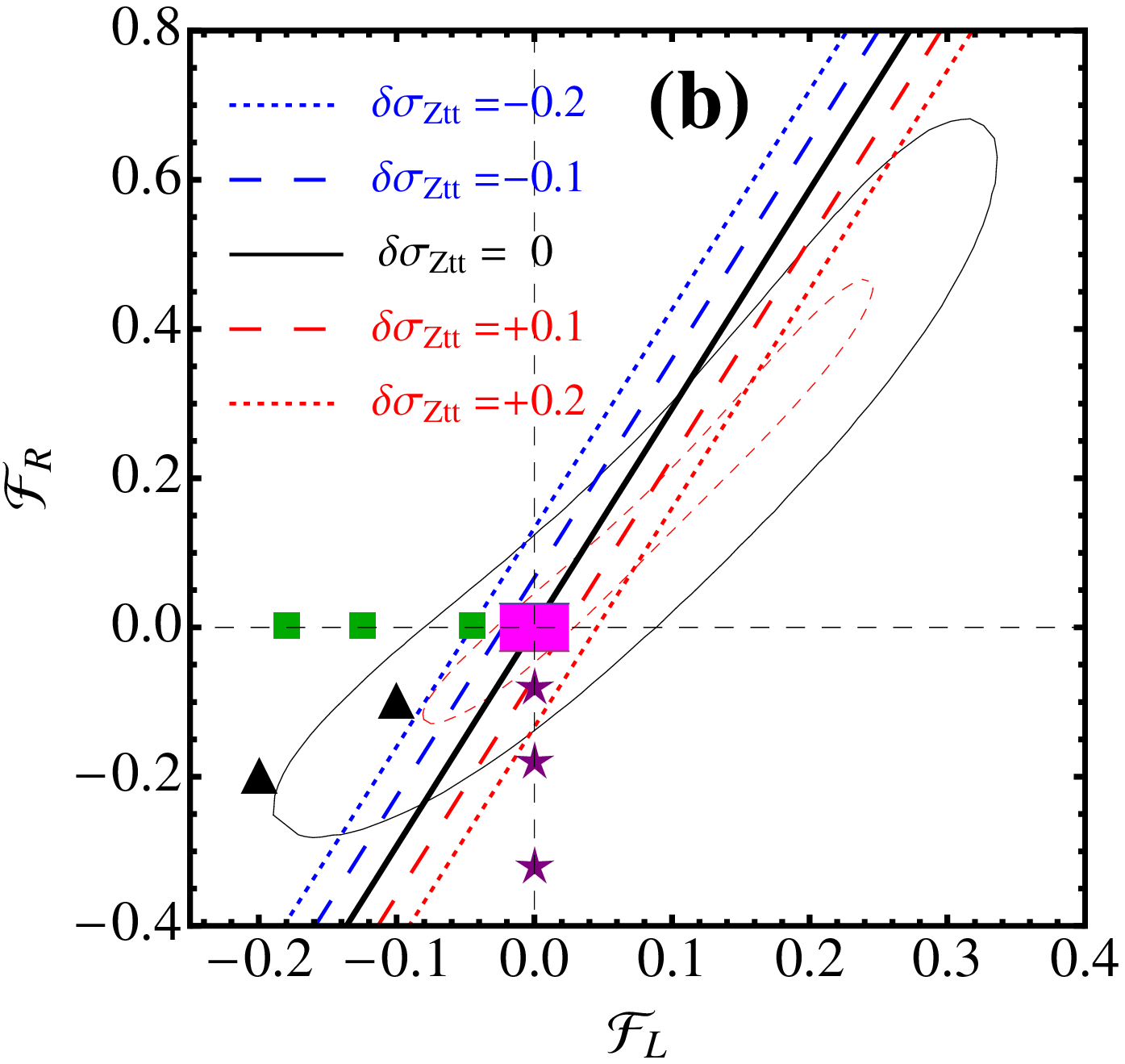}
\caption{(a): Projected $68.3\%$ C.L. bounds on the anomalous $\ztt$ coupling
from the LHC with an integrated luminosity of $300\,{\rm fb}^{-1}$(solid)
and $3000\,{\rm fb}^{-1}$(dashed) and from a LC ($\sqrt{s}=500\,{\rm GeV}$)
with an integrated luminosity of $100\,{\rm fb}^{-1}$ (magenta area).
The expected accuracy of the single top quark production cross section 
is represented by vertical bands, green (blue) for $\delta\sigma_{t}<\pm5(\pm10)\%$).
(b): Various NP model predictions, where box (star, triangle) denotes
the left-handed $t^{\prime}$ (right-handed $t^{\prime}$, sequential
fourth generation) model, respectively. See text for details. \label{fig:LCreach}}
\end{figure}

Figure~\ref{fig:LCreach} shows that new physics cannot be tested or 
excluded beyond the $2\sigma$ level without violating the bound from current 
low energy precision data discussed in Sec.~\ref{subsect:constraint}. 
But one should keep in mind that the low energy
electroweak bounds could be diluted by additional tree-level induced
operators.  Our collider simulation is based on only two effective couplings
and does not depend on four-fermion operators, making it less model dependent. 
If no other operators are present, a significant increase of linear collider
energy and improvement of detector acceptance would be needed in order to 
test new physics via correlations between the $\wtb$ and $\ztt$ couplings.

%
%
\section{Summary and Conclusions}

In this paper we investigate correlations among the values of 
the $Wtb$, $Zb{\bar b}$, and $Zt{\bar t}$ couplings.
Two main contributions are made.  First, we use a model-independent
effective Lagrangian approach to parametrize the possible effects of new physics beyond
the SM in terms of higher dimensional operators constructed from SM fields.   We
demonstrate that the $SU(2)_{L}\times U(1)_{Y}$ symmetry of the SM yields correlations
among the possible deviations of the $Zb_{L}b_{L}$, $Zt_{L}t_{L}$,
and $Wt_{L}b_{L}$ couplings.   Imposing the stringent experimental bound on $Zb_{L}b_{L}$
from data on $R_{b}$ and $A_{FB}^{b}$,
we show that a unique prediction follows on the $Zt_{L}t_{L}$ and
$Wt_{L}b_{L}$ couplings.   The prediction is independent of the underlying new
physics at the electroweak scale.   We use existing experimental constraints to show
that deviations of the $Wtb$ and $Zt{\bar{t}}$ couplings from their SM values can depend
on only two parameters, $\fl$ and $\fr$, and we present the allowed
ranges of these parameters for different new physics cutoff scales
and Higgs boson masses.

In the second contribution in this paper, we study the prospects for determining
$\fl$ and $\fr$ with LHC data.  We focus on single top
quark and $\ztt$ associated production. Deviations from the SM of
the single top quark total and differential cross sections depend
only on $\fl$, whereas the cross section for $\ztt$ associated production
is influenced by both $\fl$ and $\fr$.   We perform a detailed Monte Carlo simulation
of the $\ztt$ process, including the effects of experimental cuts.   Among
observables in $\ztt$ that  could be sensitive to the presence of $\fl$ and $\fr$,
we examine correlations between the top quark spin and the charged lepton
from $t\to bW^{+}(\to\ell^{+}\nu)$.   We find that experimental cuts appear to
leave a sample of effectively unpolarized top quarks, a disappointing conclusion
that does not appear to have been established previously.  Among the final state
distributions in $\ztt$ associated production, we confirm that the opening angle
between the two charged leptons from the $Z$ boson decay,
$\Delta\phi(\ell^{\prime+}\ell^{\prime-})$, seems to be useful for limiting the
couplings $\fl$ and $\fr$.   The principal results of our analysis are presented in
Figs.~\ref{fig:lhcreach} and~\ref{fig:LCreach} as a set of two-dimensional
plots of the deviation from the SM of the associated production cross
section $\delta\sigma_{\ztt}$ versus the deviation of the single
top quark cross section $\delta\sigma_{t}$ for integrated luminosities
of $300\,{\rm fb}^{-1}$ and $3000\,{\rm fb}^{-1}$ at the LHC. These
expectations are contrasted with estimates for a high energy electron-positron
linear collider (LC).  A comparison is also shown with predictions based
on a few recent models of new physics including a top-prime model~\cite{Dobrescu:2009vz}, 
a right-handed $t'$ model~\cite{Chivukula:1998wd}, and a model with sequential fourth generation
quarks that mix with the third generation~\cite{Kribs:2007nz}.  Our analysis suggests that about
$3000\,{\rm fb}^{-1}$ at the LHC would be needed to improve model
independent constraints on  $\fl$ and $\fr$ to the same level achievable at
a 500 GeV LC with $100\,{\rm fb}^{-1}$ of integrated luminosity.  

We also point out that the pattern of deviations predicted by
the electroweak symmetry could allow us to separate the effect of
the CKM element $|V_{tb}|$ from that of the gauge coupling $g_{Wtb}$, 
which is also present in the $\ztt$ matrix element. 
Measurements of the $\ztt$ coupling could then be used as a constraint
in analyses of data on single top quark production in order to
determine $\delta V_{tb}$. A non-zero $\delta V_{tb}$ would 
indicate non-unitarity of the $3 \times 3$ CKM matrix and 
be a clear signal for physics beyond the SM.

\begin{acknowledgments}
Q.\ H.\ C. is grateful to C.-P. Yuan for interesting and useful discussions.
E.\ L.\ B.\ and I.\ L.\ are supported by the U.~S.\ Department of Energy under
Contract No.\ DE-AC02-06CH11357. 
Q.\ H.\ C.\ is supported in part by the Argonne National Laboratory
and University of Chicago Joint Theory Institute (JTI) Grant 03921-07-137,
and by the U.S.~Department of Energy under Grants No.~DE-AC02-06CH11357
and No.~DE-FG02-90ER40560. I.\ L.\  also acknowledges the hospitality
of the Aspen Center for Physics where part of this work was completed.

\bibliographystyle{apsrev}
\bibliography{reference}

\end{acknowledgments}

\end{document}